\documentclass[review,12pt]{elsarticle}

\usepackage{caption,subcaption}
\usepackage{amssymb}
\usepackage{amsmath}
\usepackage{cleveref}
\usepackage{url}
\usepackage{natbib}

\def\ppa#1#2{\frac{\partial#1}{\partial#2}}

\def\bvec#1{\vec{\bf #1}}
\def\be{\begin{equation}}
\def\ee{\end{equation}}

\journal{arXiv}

\begin{document}

\begin{frontmatter}

	\title{High-order meshless global stability analysis of Taylor-Couette flows in complex domains}

	\author{Akash Unnikrishnan$^{a,}$}
	\author{Vinod Narayanan$^{a}$}
	\author{Surya Pratap Vanka$^{b}$\fnref{Corresponding Author}}
	\address{a Department of Mechanical Engineering,\\
		Indian Institute of Technology Gandhinagar,		Gandhinagar, Gujarat 382055, India}
	\address{b Department of Mechanical Science and Engineering,\\
		University of Illinois at Urbana-Champaign, Urbana, Illinois 61801, USA}
	\fntext[Author to whom correspondence should be addressed]{\vspace{0.3cm}Corresponding author email address: \url{spvanka@illinois.edu}}

\begin{abstract}
Recently, meshless methods have become popular in numerically solving partial differential equations and have been employed to solve equations governing fluid flows, heat transfer, and species transport. In the present study, a numerical solver is developed employing the meshless framework to efficiently compute the hydrodynamic stability of fluid flows in complex geometries. The developed method is tested on two cases of Taylor-Couette flows. The concentric case represents the parallel flow assumption incorporated in the Orr-Sommerfeld model and the eccentric Taylor-Couette flow incorporates a non-parallel base flow with separation bubbles. The method was validated against earlier works by \citet{marcus1984simulation}, \citet{oikawa1989}, \citet{leclercq2013temporal}, and \citet{mittal2014finite}. The results for the two cases and the effectiveness of the method are discussed in detail. The method is then applied to Taylor-Couette flow in an elliptical enclosure and the stability of the flow is investigated.
\end{abstract}

\begin{keyword}
Global stability analysis, meshless methods, radial basis functions, Taylor-Couette instability, elliptical enclosure
\end{keyword}

\end{frontmatter}


\section{Introduction}
\label{sec:intro}
In fluid dynamics, hydrodynamic stability stands as a fundamental pillar for understanding and predicting the behavior of fluid motion. Beyond its theoretical significance, hydrodynamic stability holds practical importance across various disciplines, from engineering to meteorology and many more. It helps in understanding the transition from well-ordered laminar flows to chaotic turbulent flows. The first step in the transition process is the amplification of small disturbances which is well captured by the Linear Stability Theory (LST). The literature on hydrodynamic stability is rich. The fundamental aspects of hydrodynamic stability were identified and studied in the 19th century by \citet{helmholtz1868xliii}, \citet{kelvin1871stratified}, \citet{reynolds1883iii}, and \citet{rayleigh1882investigation}. Through his classical pipe flow experiments, Reynolds studied the transition process and showed that laminar flow breaks down and turbulence ensues when the Reynolds number exceeds a critical value. Theoretical and experimental investigations of hydrodynamic stability became prominent after the formulation of the Orr-Sommerfeld equations (\citet{drazin1961discontinuous}). The works of \citet{heisenberg1923absolute}, \citet{tollmien1936general}, \citet{schlichting1949boundary}, \citet{schubauer1947laminar}, \citet{squire1933stability} are seminal in this area.

While most of the work was concentrated on boundary layers and channel flows,  some noted works on different geometries were also reported in the literature. 
\citet{taylor1923viii} studied the stability of flow between two concentric rotating cylinders and approximated the disturbance with Bessel functions and Fourier modes. The study revealed the mechanism of the formation of Taylor cells in multiple configurations of co-rotating, and counter-rotating cylinders. The work discussed the formation of a spiral form of Taylor cells when the base flow was not restricted to two dimensions. 

The classical Linear Stability Theory which led to the Orr-Sommerfeld equations assumes that the base flow is parallel. This assumption holds good for channel flows.  It was shown through a series of works by \citet{gaster1974effects}, \citet{saric1975nonparallel}, \citet{fasel1990non}, and \citet{bertolotti1992linear} that nonparallel effects in boundary layers are very small. \citet{gaster1974effects} studied the stability of flat plate boundary layer using asymptotic series solutions and iterative solution methods. \citet{saric1975nonparallel} analyzed non-parallel effects of boundary layer using the method of multiple scales. \citet{fasel1990non} did a numerical integration of Navier Stokes equations to study spatially evolving waves in the flat plate boundary layer and argued that the effects of the non-parallel nature of the flow on the growth of instabilities are negligible. \citet{bertolotti1992linear} studied the linear and nonlinear instability of the growing boundary layer and found that the effect of nonlinearity is dominant over the nonparallel nature of the boundary layer and accounts for the discrepancy with experimental measurements. Flows with significant nonparallel effects are studied using Global Stability Theory (GST) (\citet{theofilis2011}, \citet{bhoraniya2017global}, and \citet{bhoraniya2018global}). 

The flow between two concentric cylinders with the inner/outer cylinder rotating, is one-dimensional and parallel because the tangential velocity depends only on the radius. Due to the absence of the radial velocity component, the base flow and the perturbation equations become quite simplified. However, if the inner cylinder is placed eccentrically, the gap between the inner and outer cylinders varies with angular position and the resulting laminar base flow is no longer axisymmetric. Both radial and tangential velocity components depend on the radius as well as the angle, making the flow non-parallel. For such cases, a bi-global stability analysis is required. Also, the disturbances cannot be assumed to be Fourier modes as in the case of parallel flows.

Typical global stability can result in a very large eigenvalue problem with operations on matrices of size 40,000 or more. The commonly used methods are spectral methods (e.g. \citet{marcus1984simulation}, \citet{karniadakis1992three}, \citet{thompson1996three}, \citet{barkley1999stability}, \citet{bhoraniya2017global}, \citet{bhoraniya2018global}, \citet{dandelia2022optimal}), finite difference methods (e.g. \citet{ramanan1994linear}), asymptotic analyses (e.g. \citet{garg1972linear}, \citet{yiantsios1988linear}), and finite element methods (e.g. \citet{saraph1979stability}, \citet{mittal2014finite}). Spectral methods are very accurate but are limited to geometries that can be represented by spectral expansions. Further, they are restricted to the placement of grid points at specified locations such as at Gauss-Lobatto points of Chebychev polynomials. Global stability analysis of a flow in a complex domain becomes difficult because of two reasons. First, computing the base flow is expensive. Second, the linear stability equations and boundary conditions are not straightforward. Complicated coordinate transformations are required to obtain the linear stability equations and the homogeneous boundary conditions. A subsequent complexity is the false prediction of instabilities resulting from the amplification of numerical diffusion and spurious oscillations (\citet{renardy1986linear}, \citet{malik1987linear}, \citet{michalcova2020numerical}) resulting from the improper choice of numerical methods. For example, finite-difference methods in curvilinear coordinate systems or finite-element methods are relatively low order and require fine grids to obtain the desired accuracy for stability analyses in complex domains. Therefore, the numerical method should first be carefully validated to ensure the accuracy of discretization and avoid spurious growth of numerical errors. For complex geometries, spectral methods require coordinate transformations which makes the resulting set of equations complex and cumbersome to solve. The spectral element method (\citet{patera1984spectral}, \citet{lee2009spectral}) discretizes the domain in arbitrary quadrilateral or hexahedral elements and employs separate spectral expansions inside each element, thus combining the geometric flexibility of the finite element method with the accuracy of spectral methods. The spectral element method has been applied to study flow stability by \citet{hill2006legendre}, and \citet{yin2016linear}.

\citet{marcus1984simulation} developed a pseudospectral method to numerically solve Navier Stokes equations using Green's functions to remove time-splitting errors. This method was used to study the local stability of the Taylor-Couette flow. Stability of different variations of  Taylor-Couette flow was studied by \citet{jenny2007primary}, \citet{white2000viscous}, \citet{eagles1997stability}, \citet{kedia1998numerical}, and \citet{al2002effect}, where heating and gravitational effects were considered. \citet{wannier1950contribution} did a comparative study of the Reynolds equation and Stokes equation of the eccentric Taylor-Couette model. He gave analytical expressions for limiting cases of eccentric Taylor-Couette flow. \citet{vohr1968} conducted experiments on eccentric Taylor-Couette flow to determine the critical speed ratio for various eccentricity values. \citet{oikawa1989} used a Chebychev-Fourier basis to solve linearised Navier Stokes equations in a bipolar coordinate system to analyze the stability of eccentric Taylor-Couette flow. \citet{mittal2014finite} developed a finite element framework for performing global linear stability analysis of two-dimensional flows. \citet{leclercq2013temporal,leclercq2014absolute} have worked on absolute and temporal instabilities of Taylor-Couette Poiseuille flows using linear stability analysis in a bipolar coordinate system.

Numerical investigations of global stability have so far been limited to simple geometries due to computational limitations. 
In contrast with mesh-based discretization of Navier-Stokes equations, recently there has been significant interest in meshless methods to discretize partial differential equations.  Meshless methods discretize the complex domain with scattered points and interpolate data with special functions. Since the early works by \citet{hardy1971multiquadric}, several different RBFs have been evaluated for accuracy, and algorithms have been developed to solve partial differential equations (\citet{KANSA1990127}, \citet{KANSA1990147}, \citet{KANSA2000123}, \citet{Bayona2017}). 
Here, the equations are solved in a Cartesian framework with the order of accuracy controlled by the degree of appended polynomials.

\begin{figure}
    \centering
    \begin{subfigure}[b]{0.32\textwidth}
        \includegraphics[width = \textwidth]{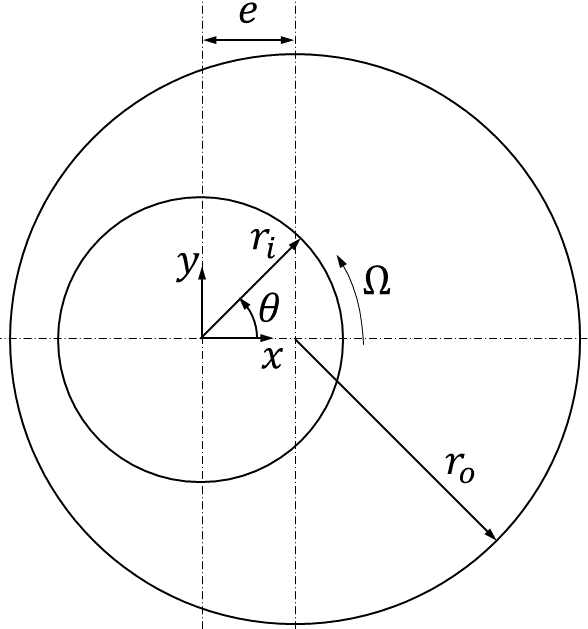}
        \caption{}
        \label{schematic_a}
    \end{subfigure}
    \hfill
    \begin{subfigure}[b]{0.64\textwidth}
        \includegraphics[width = \textwidth]{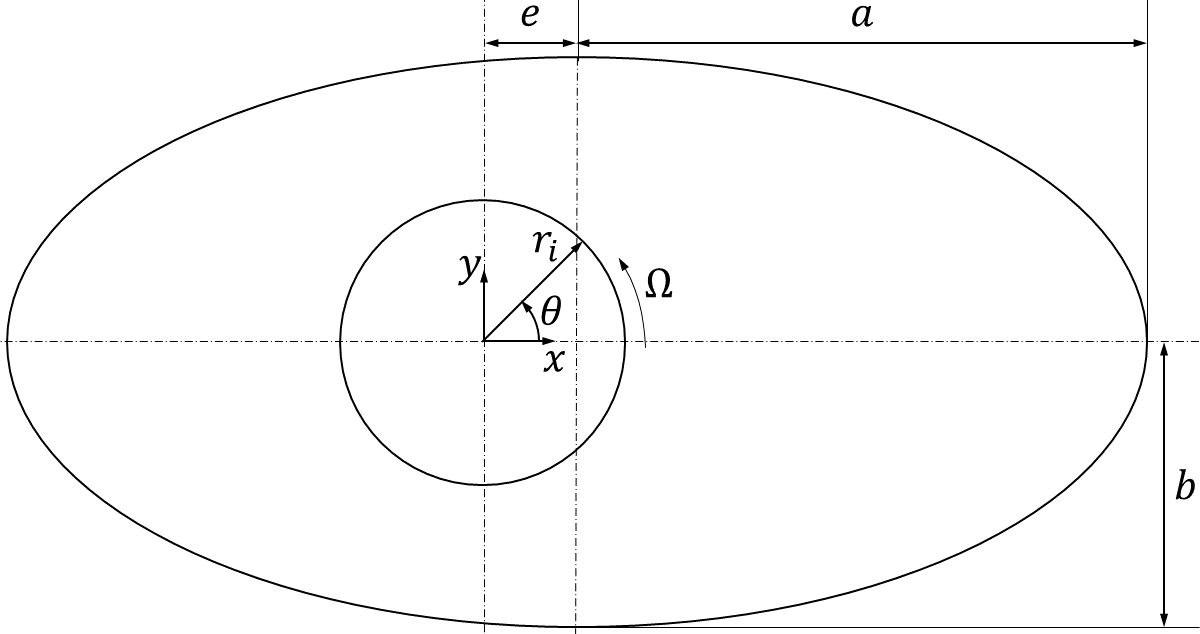}
        \caption{}
        \label{schematic_b}
    \end{subfigure}
    \caption{Schematic of geometries studied: a) Taylor-Couette flows between concentric and eccentric circular cylinders, and b) Taylor-Couette flow in an elliptical outer enclosure.}
    \label{fig:schematic}
\end{figure}

In the present paper, we describe an algorithm to perform global stability analyses in arbitrary complex domains using meshless methods. Our approach extends recent works (\citet{Shantanu2016}, \citet{shahane2021high}, \citet{radhakrishnan2021non}, \citet{bartwal2021application}, \citet{unnikrishnan2024taylor}) to solve heat conduction and fluid flow in complex domains using high order accurate polyharmonic spline radial basis functions (PHS-RBF). First, a two-dimensional base flow is generated by solving the incompressible Navier-Stokes equations in a complex domain using the high-order meshless algorithm. \Cref{fig:schematic} describes the two flow configurations studied: Taylor-Couette flow between two circular cylinders and between a circular cylinder and an ellipse. The base flow is generated at a Reynolds number close to the expected transition state.  The Navier-Stokes equations are then perturbed and the eigenvalue problem for the linearized three-dimensional N-S equations for perturbations is formulated.  The novelty of the present work is that all derivatives appearing in the perturbed Navier-Stokes equations are also computed by differentiating the same radial basis function interpolations.  Thus, the method reduces complexities related to the coordinate transformations and/or domain transformations. A sparse matrix system, as in the case of a finite difference scheme, is then solved but with no underlying grid.  Recently, \citet{chu2024mesh} have also reported a similar approach to study the two-dimensional hydrodynamic stability of complex flows. They considered linear stability analysis and a resolvent analysis of the Blassius-boundary layer, flow over a circular cylinder, and a transonic turbulent jet. Their study, however, considered only two-dimensional perturbations as the flows first undergo two-dimensional instabilities. Our algorithm permits three-dimensional perturbations on two-dimensional steady base flows.  

We first describe in \cref{sec:governing_eqns} the governing equations and their perturbation form. In \cref{sec:numerical}, the concept of meshless solution of partial differential equations is described. \Cref{sec:validation} demonstrates the algorithm by validating it in two canonical Taylor-Couette flows, the concentric and eccentrically placed inner rotating cylinder inside an outer circular enclosure.  In \cref{sec:results}, we present results of the instability of the flow in an elliptical enclosure with a rotating inner circular cylinder. The base flow, obtained in a previous study (\citet{unnikrishnan2022shear}) is presented for a $Re$ of 48.5 to which perturbations are applied. The results show the most amplified perturbations and the Reynolds number at which they first appear to grow. \Cref{sec:conclusions}, provides a summary of current findings.
 
 
 

\section{Governing equations}

\label{sec:governing_eqns}
  
We begin with continuity and momentum equations in Cartesian coordinates. The velocity is normalized with the tangential velocity ($r_i\Omega$) of the inner cylinder and the length scale is normalized with the radius of the inner cylinder ($r_i$). $\Omega$ is the angular velocity of the cylinder. The Reynolds number is defined as,
\begin{equation}
    Re=\frac{r_i^2\Omega}{\nu}
\end{equation}
where $\nu$ is the kinematic viscosity of the fluid. 
The non-dimensionalised continuity and momentum equations for incompressible flow are,
\begin{equation}
    \nabla. \bvec V =0
\end{equation}
\begin{equation}
   \ppa {\bvec V} t + (\bvec V.\nabla){\bvec V}   =  -\nabla P + \frac {1} {Re} \nabla^2 {\bvec V}
\end{equation}
where $ \bvec V$ is the instantaneous velocity vector.
 
In the case of two concentric cylinders, the base flow is parallel, with only a single component of velocity in the $\theta$ direction and zero radial velocity.
The instantaneous velocity is first written as a sum of a base flow and perturbations.
Let the field variables be of the form $\boldsymbol{q} = \boldsymbol{Q} + \boldsymbol{q}'$; where $\boldsymbol{q}$ is the instantaneous flow field $[u, v, w, p]^T$ and $\boldsymbol{Q}$ is the base flow field $\boldsymbol{Q}=[U,V,W,P]^T$. The equations for the mean flow are then subtracted from the respective equations for the instantaneous velocity field. The resulting equations are linearised, neglecting the products of two perturbations. The linear stability equations written in Cartesian coordinates are

        \label{eqn:z_momentum}


     Continuity equation
    \begin{equation}
        \frac{\partial u'}{\partial x}+\frac{\partial v'}{\partial y}+ \frac{\partial w'}{\partial z} =0
        \label{eqn:pert_continuity}
    \end{equation}
Momentum equations
\begin{subequations}
    \begin{equation}
        \frac{\partial u'}{\partial t} + u'\frac{\partial U}{\partial x} + U\frac{\partial u'}{\partial x} + v'\frac{\partial U}{\partial y} + V\frac{\partial u'}{\partial y} + w'\frac{\partial U}{\partial z}= -\frac{\partial p'}{\partial x}+ \frac{1}{Re}(\frac{\partial^2 u'}{\partial {x}^2} +\frac{\partial^2 u'}{\partial {y}^2} +\frac{\partial^2 u'}{\partial {z}^2}) 
    \end{equation}
    \begin{equation}
        \frac{\partial v'}{\partial t} + u'\frac{\partial V}{\partial x} + U\frac{\partial v'}{\partial x} + v'\frac{\partial V}{\partial y} + V\frac{\partial v'}{\partial y} + w'\frac{\partial V}{\partial z}= -\frac{\partial p'}{\partial y}+ \frac{1}{Re}(\frac{\partial^2 v'}{\partial {x}^2} +\frac{\partial^2 v'}{\partial {y}^2} +\frac{\partial^2 v'}{\partial {z}^2}) 
    \end{equation}
    \begin{equation}
        \frac{\partial w'}{\partial t} + U\frac{\partial w'}{\partial x} + V\frac{\partial w'}{\partial y} = -\frac{\partial p'}{\partial z}+ \frac{1}{Re}(\frac{\partial^2 w'}{\partial {x}^2} +\frac{\partial^2 w'}{\partial {y}^2} +\frac{\partial^2 w'}{\partial {z}^2}) 
    \end{equation}
\label{eqn:pert_mom}
\end{subequations}

We perform normal mode analysis and assume perturbations of the form
\begin{equation}
    \boldsymbol{q'} = \boldsymbol{\hat{q}}(x,y) \exp{\iota(\alpha z - \omega t)}
    \label{eqn:normal_mode}
\end{equation}
where $\alpha$ is the wavenumber.  $\omega$ is a complex number ($\omega_r+ \iota\omega_i$). The real part ($\omega_r$) is the frequency of the wave and the imaginary part ($\omega_i$) represents the temporal growth rate.
In the present study, the axial velocity component of the base flow ($W$) is zero. Further, the gradients of base velocity in the axial direction($\partial U/\partial z, \partial V/\partial z$) are also zero.
Substituting the normal mode form (\cref{eqn:normal_mode}) in \cref{eqn:pert_mom}, we obtain


\begin{subequations}
\begin{align}
        0 &=D_x \hat{u}+  D_y\hat{v} - \alpha \tilde{w} 
        \label{eqn:nm_pert_continuity}\\
        i{\omega} \hat{u} &= D_x \hat{p} + (\mathcal{L} + D_x U) \hat{u} + (D_y U) \hat{v} 
        \label{eqn:nm_pert_x_momentum}\\
        i{\omega} \hat{v} &= D_y \hat{p} + (D_x V)\hat{u} + (\mathcal{L} + D_y V) \hat{v}
        \label{eqn:nm_pert_y_momentum}\\
        i{\omega} \tilde{w} &= \alpha\hat{p} + \mathcal{L} \tilde{w}
        \label{eqn:nm_pert_z_momentum}
\end{align}
\end{subequations}

where $ \mathcal{L} \equiv \left(U D_x + V D_y \right) - \frac{1}{Re}\left(D_x^2 + D_y^2 -\alpha^2 \right)$, $D_x \equiv \frac{\partial}{\partial x}$, $D_y \equiv \frac{\partial}{\partial y}$, $\tilde{w} = i\hat{w}$, and $D$ denotes derivative in the directions mentioned by the subscript. 

This linear system of equations can be written in matrix-vector form as 
\begin{equation}
    \begin{bmatrix}
        \boldsymbol{\mathcal{L}}+\boldsymbol{D_x}U & \boldsymbol{D_y}U & \boldsymbol{0} & \boldsymbol{D_x}\\
        \boldsymbol{D_x}V &  \boldsymbol{\mathcal{L}}+\boldsymbol{D_y}V & \boldsymbol{0} & \boldsymbol{D_y}\\
        \boldsymbol{0} & \boldsymbol{0} & \boldsymbol{\mathcal{L}} & \alpha \boldsymbol{I}\\
        \boldsymbol{D_x} & \boldsymbol{D_y} & -\alpha \boldsymbol{I} & \boldsymbol{0}
    \end{bmatrix}
    \begin{bmatrix}
        \hat{u} \\ \hat{v} \\ \tilde{w} \\ \hat{p}
    \end{bmatrix}
    =i \omega 
    \begin{bmatrix}
        \boldsymbol{I} & \boldsymbol{0} & \boldsymbol{0} & \boldsymbol{0}\\
        \boldsymbol{0} & \boldsymbol{I} & \boldsymbol{0} & \boldsymbol{0} \\
        \boldsymbol{0} & \boldsymbol{0} & \boldsymbol{I} & \boldsymbol{0} \\
        \boldsymbol{0} & \boldsymbol{0} & \boldsymbol{0} & \boldsymbol{0}
    \end{bmatrix}
    \begin{bmatrix}
        \hat{u} \\ \hat{v} \\ \tilde{w} \\ \hat{p}
    \end{bmatrix}
\end{equation}
\begin{equation}
    \textbf{A}\boldsymbol{\hat{q}} = i\omega\textbf{B}\boldsymbol{\hat{q}}
    \label{eqn:bigABmatrices}
\end{equation}
where \textbf{A} and \textbf{B} are linear operators on $\boldsymbol{\hat{q}}$, and $i\omega$ is the eigenvalue corresponding to the eigenvector $\boldsymbol{\hat{q}} = [\hat{u},\hat{v},\tilde{w},\hat{p}]^T$.

At the walls, impermeability and no-slip conditions are used, which gives $u=v=w=0$. These boundary conditions can be imposed by adjusting the coefficients of the $\boldsymbol{A}$ and $\boldsymbol{B}$ matrices corresponding to the boundary points. We use pressure compatibility conditions as described in \citet{theofilis2011}. The governing equations are invariant under complex conjugation and hence we require only to investigate the positive values of $\alpha$. The present study investigates purely temporal growth (spatial growth of perturbations is neglected), which makes $\alpha$ a real number. The problem is also invariant to the axial reflection, and hence the eigenvalues are in pairs and symmetric about the imaginary axis. The temporal modes $\omega$ are sought for a given value of $(\alpha, Re, e, \eta)$, where $e$ is the eccentricity and $\eta$ is the radius ratio given by $\eta = r_i/r_o$. $\eta$ is defined only for the flow between the two circular cylinders. For the elliptical enclosure, we analyze the case with an aspect ratio of 2. The aspect ratio is defined as the ratio of the major to the minor axes of the ellipse. Computations of this nature, require varying the parameters $(\alpha, Re)$ in an $\mathcal{R}^2$ space for a given eccentricity, with separate solutions of the eigenvalue problem for each case. The critical values of $Re$ and $\alpha$ are defined when the absolute value of the mode $\omega$ is within a tolerance of 1e-4.

\section{Numerical Solution}
\label{sec:numerical}

Because of the complex geometry, a meshless method is employed to solve for the eigenvalues and eigenvectors of the big sparse matrices $\boldsymbol{A}$ and $\boldsymbol{B}$ in the \cref{eqn:bigABmatrices}. Meshless methods, as the name suggests, do not involve element or node connectivity. The method represents a complex geometry only by a set of scattered points and interpolates variables using radial basis functions (RBF). Several different radial basis functions have been used in literature, such as multiquadrics, inverse multiquadrics, Gaussian, polyharmonic splines, and others. The absence of a shape parameter makes the polyharmonic spline (PHS) function attractive over multiquadrics and Gaussian RBF. The elegance of the meshless method is that since it does not require control volumes and element connectivity, its accuracy is not adversely affected by the skewness of the grid. Further, the PHS-RBF is appended with a polynomial, which controls the order of interpolation accuracy and the spatial derivatives. For details of RBF interpolation, see [\cite{Bayona2017}, \cite{flyer2016onrole_I}, \cite{Shantanu2016}, \cite{shahane2021high}, \cite{chu2023rbf}].

\begin{figure}
    \centering
    \begin{subfigure}[b]{0.32\textwidth}
        \includegraphics[width =\textwidth]{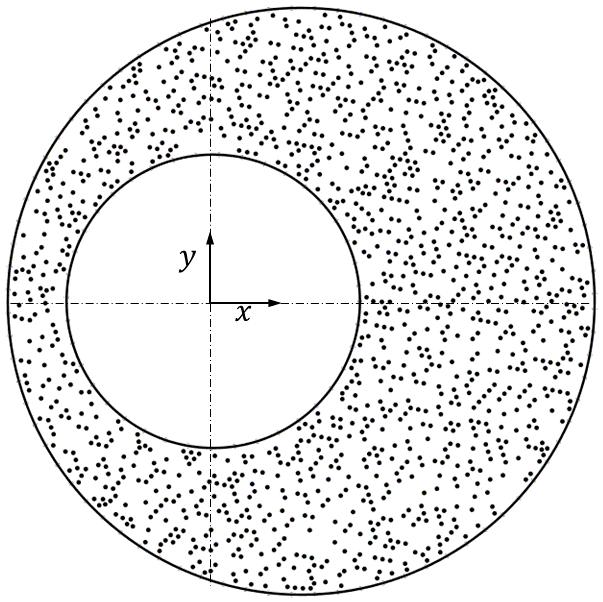}
        \caption{}
        \label{fig:poin_dist_a}
    \end{subfigure}
    \hfill
    \begin{subfigure}[b]{0.64\textwidth}
        \includegraphics[width = \textwidth]{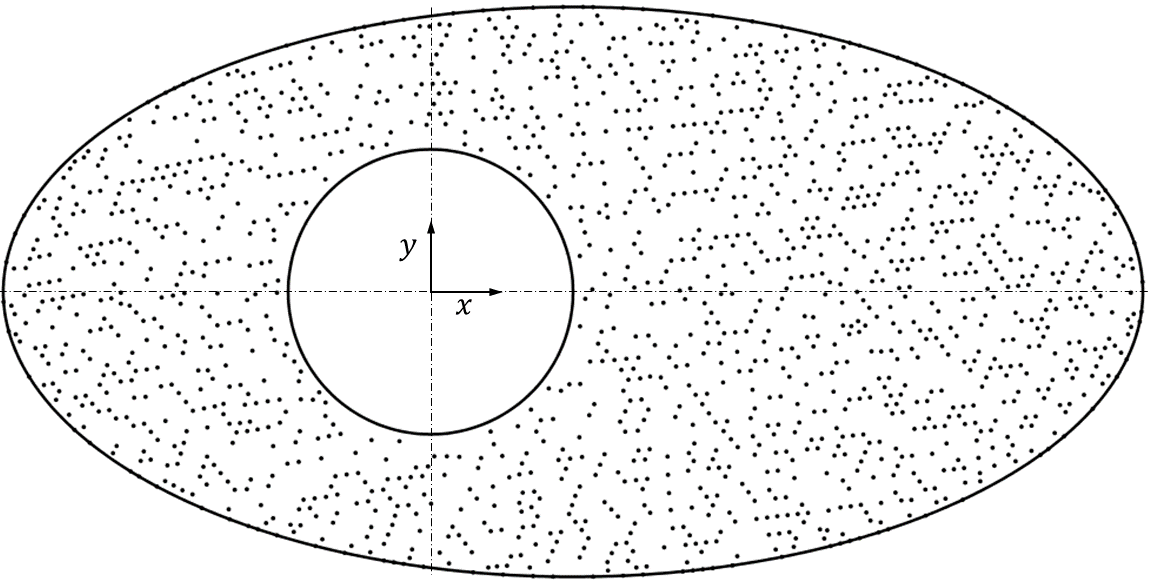}
        \caption{}
        \label{fig:poin_dist_b}
    \end{subfigure}
    \caption{Sample point distributions for a) eccentric Taylor-Couette flow, and b) Taylor-Couette flow in an elliptical outer enclosure.}
    \label{fig:point_dist}
\end{figure}

\Cref{fig:point_dist} gives an example of a scattered point distribution for the eccentric Taylor-Couette flow and the elliptical enclosure. The first step in the meshless method is to interpolate a variable between the scattered points. A cloud-based interpolation is used in which any variable at a point is interpolated over a cloud of neighboring points using the Polyharmonic Spline - Radial Basis Functions (PHS-RBF).
The required derivative operators $D_x$ and  $D_y$ and the Laplacian operator ($D_x^2  + D_y^2$)  are calculated by differentiating the interpolating function at the scattered points.  An arbitrary variable $s(\boldsymbol{x})$ is interpolated as 
\begin{equation}
    s(\boldsymbol{x}) = \sum_{i=1}^{q} \lambda_i \phi (||\boldsymbol{x} - \boldsymbol{x_i}||_2) + \sum_{i=1}^{m} \gamma_i P_i (\boldsymbol{x})
    \label{Eq:RBF_interp}
\end{equation}
where, $\phi(r)=r^{2a+1},\hspace{0.1cm} a \in \mathbb{N}$ is the PHS-RBF, $m$ is the number of monomials ($P_i$) upto a maximum degree of $l$ and $(\lambda_i, \gamma_i)$ are $q+m$ coefficients. We use $a =1$ and $q$ equations are obtained by collocating \cref{Eq:RBF_interp} over the $q$ cloud points. The $m$ additional equations required to close the linear system are imposed as constraints on the polynomials (\cite{flyer2016onrole_I}):  
    \begin{equation}
        \sum_{i=1}^{q} \lambda_i P_j(\boldsymbol{x_i}) =0 \hspace{0.5cm} \text{for } 1 \leq j \leq m
    \label{Eq:RBF_constraint}
    \end{equation}
  In a matrix-vector form, we can write these equations as
    
    \begin{equation}
        \begin{bmatrix}
        \boldsymbol{\Phi} & \boldsymbol{P}  \\
        \boldsymbol{P}^T & \boldsymbol{0} \\
        \end{bmatrix}
        \begin{bmatrix}
        \boldsymbol{\lambda}  \\
        \boldsymbol{\gamma} \\
        \end{bmatrix} =
        \begin{bmatrix}
        \boldsymbol{C}
        \end{bmatrix}
        \begin{bmatrix}
        \boldsymbol{\lambda}  \\
        \boldsymbol{\gamma} \\
        \end{bmatrix} =
        \begin{bmatrix}
        \boldsymbol{s}  \\
        \boldsymbol{0} \\
        \end{bmatrix}
        \label{Eq:RBF_interp_mat_vec}
    \end{equation}
    where, transpose is denoted by the superscript $T$, $\boldsymbol{\lambda} = [\lambda_1,...,\lambda_q]^T$, $\boldsymbol{\gamma} = [\gamma_1,...,\gamma_m]^T$, $\boldsymbol{s} = [s(\boldsymbol{x_1}),...,s(\boldsymbol{x_q})]^T$ and $\boldsymbol{0}$ is the vector of zeros. Dimensions of the submatrices  $\boldsymbol{\Phi}$ and $\boldsymbol{P}$ are $q\times q$ and $q\times m$ respectively.

    For a two-dimensional problem with an appended polynomial of degree 2, there are $m=6$ polynomial terms, given as $[1, x, y, x^2, xy, y^2]$. For our computations, we have used appended polynomials of degree 5. With this degree of appended polynomial, the first derivative has an order of accuracy of 5 and the second derivative has an order of accuracy of 4. The differential operators ($\mathcal{D}$) can be obtained by differentiating the RBF and the polynomials. 
    \begin{equation}
        \mathcal{D} [s(\boldsymbol{x})] = \sum_{i=1}^{q} \lambda_i \mathcal{D} [\phi (\boldsymbol{||\boldsymbol{x} - \boldsymbol{x_i}||_2})] + \sum_{i=1}^{m} \gamma_i \mathcal{D}[P_i (\boldsymbol{x})]
        \label{Eq:RBF_interp_L}
    \end{equation}
    \Cref{Eq:RBF_interp_L} applied to all the points in the cloud leads to a rectangular matrix-vector system given by \cref{Eq:RBF_interp_mat_vec_L}.
    \begin{equation}
        \mathcal{D}[\boldsymbol{s}] =
        \begin{bmatrix}
        \mathcal{D}[\boldsymbol{\Phi}] & \mathcal{D}[\boldsymbol{P}]  \\
        \end{bmatrix}
        \begin{bmatrix}
        \boldsymbol{\lambda}  \\
        \boldsymbol{\gamma} \\
        \end{bmatrix}
        \label{Eq:RBF_interp_mat_vec_L}
    \end{equation}
    where, $\mathcal{D}[\boldsymbol{\Phi}]$ and $\mathcal{D}[\boldsymbol{P}]$ are matrices of sizes $q\times q$ and $q\times m$ respectively. Substituting \cref{Eq:RBF_interp_mat_vec} in \cref{Eq:RBF_interp_mat_vec_L} results in:
    \begin{equation}
        \begin{aligned}
        \mathcal{D}[\boldsymbol{s}] &=
        \left(\begin{bmatrix}
        \mathcal{D}[\boldsymbol{\Phi}] & \mathcal{D}[\boldsymbol{P}]  \\
        \end{bmatrix}
        \begin{bmatrix}
        \boldsymbol{C}
        \end{bmatrix} ^{-1}\right)
        \begin{bmatrix}
        \boldsymbol{s}  \\
        \boldsymbol{0} \\
        \end{bmatrix}
        =
        \begin{bmatrix}
        \boldsymbol{G}
        \end{bmatrix}
        \begin{bmatrix}
        \boldsymbol{s}  \\
        \boldsymbol{0} \\
        \end{bmatrix}\\
        &=
        \begin{bmatrix}
        \boldsymbol{G_1} & \boldsymbol{G_2}
        \end{bmatrix}
        \begin{bmatrix}
        \boldsymbol{s}  \\
        \boldsymbol{0} \\
        \end{bmatrix}
        = [\boldsymbol{G_1}] [\boldsymbol{s}] + [\boldsymbol{G_2}] [\boldsymbol{0}]
        = [\boldsymbol{G_1}] [\boldsymbol{s}]
        \end{aligned}
        \label{Eq:RBF_interp_mat_vec_L_solve}
    \end{equation}
    From \cref{Eq:RBF_interp_mat_vec_L_solve}, any differential operator $\mathcal{D}$ is approximated by a matrix $\boldsymbol{G_1}$ that multiplies the discrete values of the variable, $s$. The gradient ($\boldsymbol{D_x}$ and $\boldsymbol{D_y}$) and Laplacian ($\boldsymbol{D_x}^2 + \boldsymbol{D_y}^2$) operators, can be generated by finding the corresponding matrix operators using a similar procedure. 
    By substituting the expressions for the operators in \cref{eqn:nm_pert_continuity,eqn:nm_pert_x_momentum,eqn:nm_pert_y_momentum,eqn:nm_pert_z_momentum} we can now generate the complete $\boldsymbol{A}$ and $\boldsymbol{B}$ matrices for a given value of wave number $\alpha$, and Reynolds number $Re$.
    Further details of the meshless method can be found in \citet{Shantanu2016}, \citet{shahane2021high}, \citet{bartwal2021application}, \citet{radhakrishnan2021non}, \citet{wang2020weighted}, and \citet{chu2023rbf}.

\subsection{Computation of Eigenvalues}
The matrices $\boldsymbol{A}$ and $\boldsymbol{B}$ are real and are large sparse matrices. It is computationally expensive to directly solve for the entire spectrum of eigenvalues and eigenvectors. In the modal stability analysis, the mode with the largest imaginary value is most important, since it represents the temporal growth rate of the mode. Hence, the few of the least stable modes and their corresponding eigenvectors are sought. Arnoldi's algorithm (\citet{arnoldi1951principle}), with a spectral transformation as in \cref{eqn:arnoldi_shift_invert}, is an iterative strategy to compute a few eigenvalues and eigenvectors around the shift parameter $\lambda$.
\begin{equation}
    (\boldsymbol{A}-\lambda \boldsymbol{B})^{-1} \boldsymbol{B} \psi = \mu \psi
    \label{eqn:arnoldi_shift_invert}
\end{equation}
where  $\mu = \frac{1}{i \omega - \lambda}$, are the eigenvalues of the modified problem. The eigenvalues of the matrix $(\boldsymbol{A}-\lambda \boldsymbol{B})^{-1} \boldsymbol{B}$ with the largest magnitude correspond to the eigenvalues of the general eigenvalue problem $\boldsymbol{A} \boldsymbol{q} = i\omega \boldsymbol{B} \boldsymbol{q}$, close to the shift value, $\lambda$. These eigenvalues ($\mu$) are easy to compute by the Krylov method. The Krylov subspace may be computed by successive resolution of the linear system with matrix $(\boldsymbol{A}-\lambda \boldsymbol{B})$, using LU decomposition. The full spectrum of eigenmodes is computed in this small subspace to approximate the solution to the original general eigenvalue problem. We follow the approach mentioned in \cite{bhoraniya2017global,bhoraniya2018global} to extract the eigenvalues and eigenvectors.

\section{Validation for concentric and eccentric Taylor-Couette flows}
\label{sec:validation}
\begin{figure}
    \centering
    \begin{subfigure}[b]{0.31 \textwidth}
        \includegraphics[height=43mm]{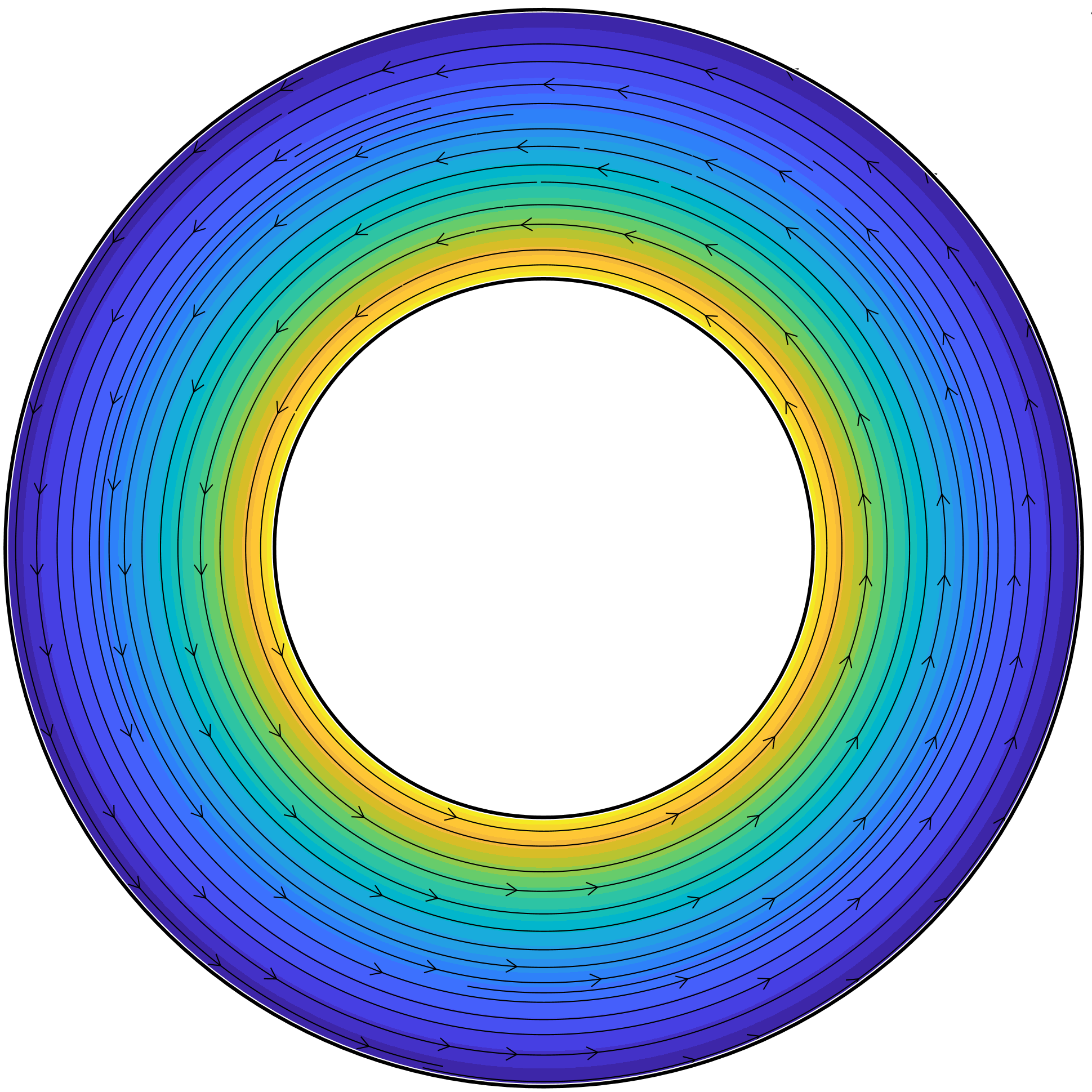}
        \caption{}
        \label{fig:base flow_A}
    \end{subfigure}
    \begin{subfigure}[b]{0.31 \textwidth}
        \includegraphics[height=43mm]{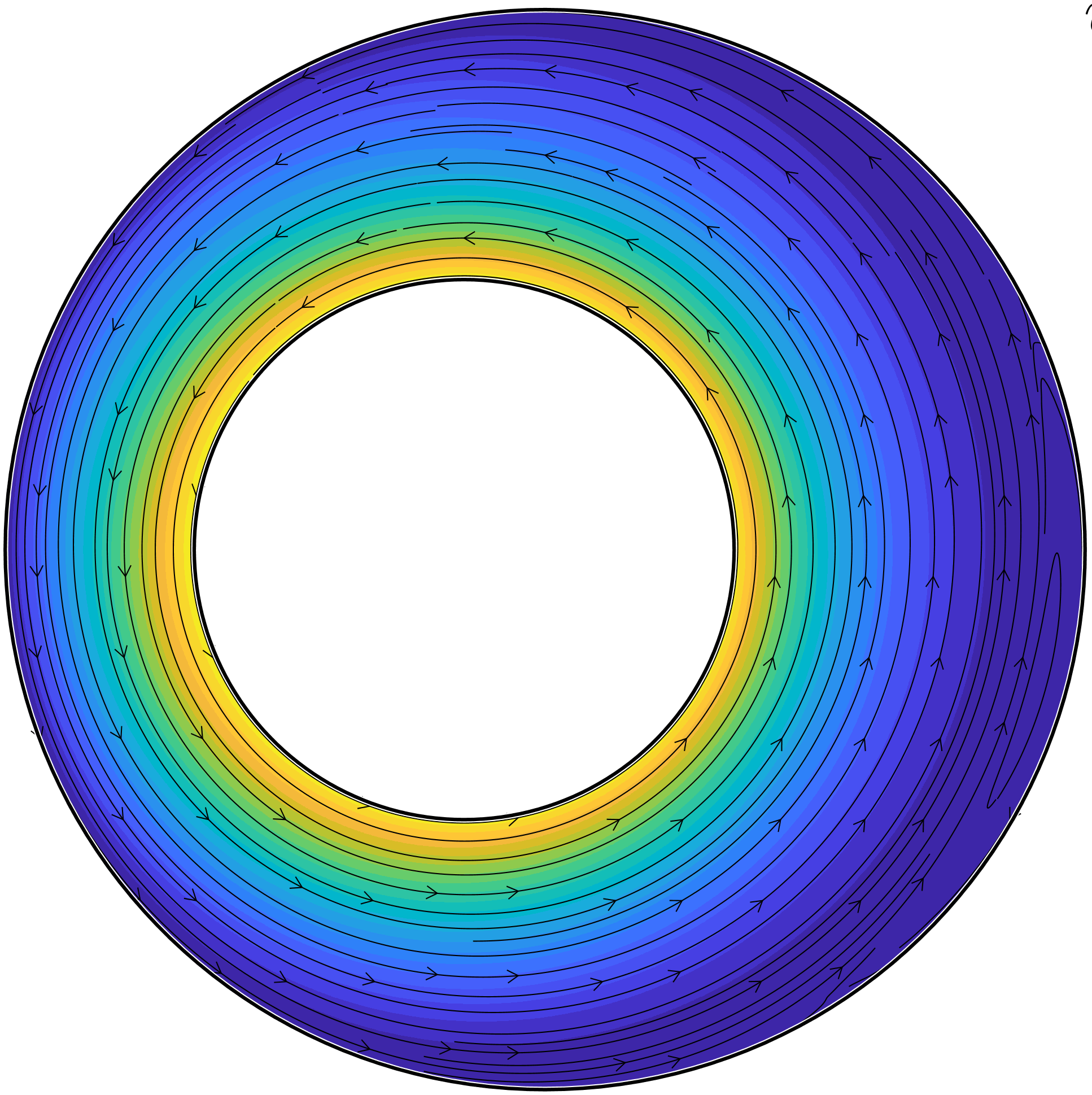}
        \caption{}
        \label{fig:base flow_B}
    \end{subfigure}
    \begin{subfigure}[b]{0.35 \textwidth}
        \includegraphics[height=43mm]{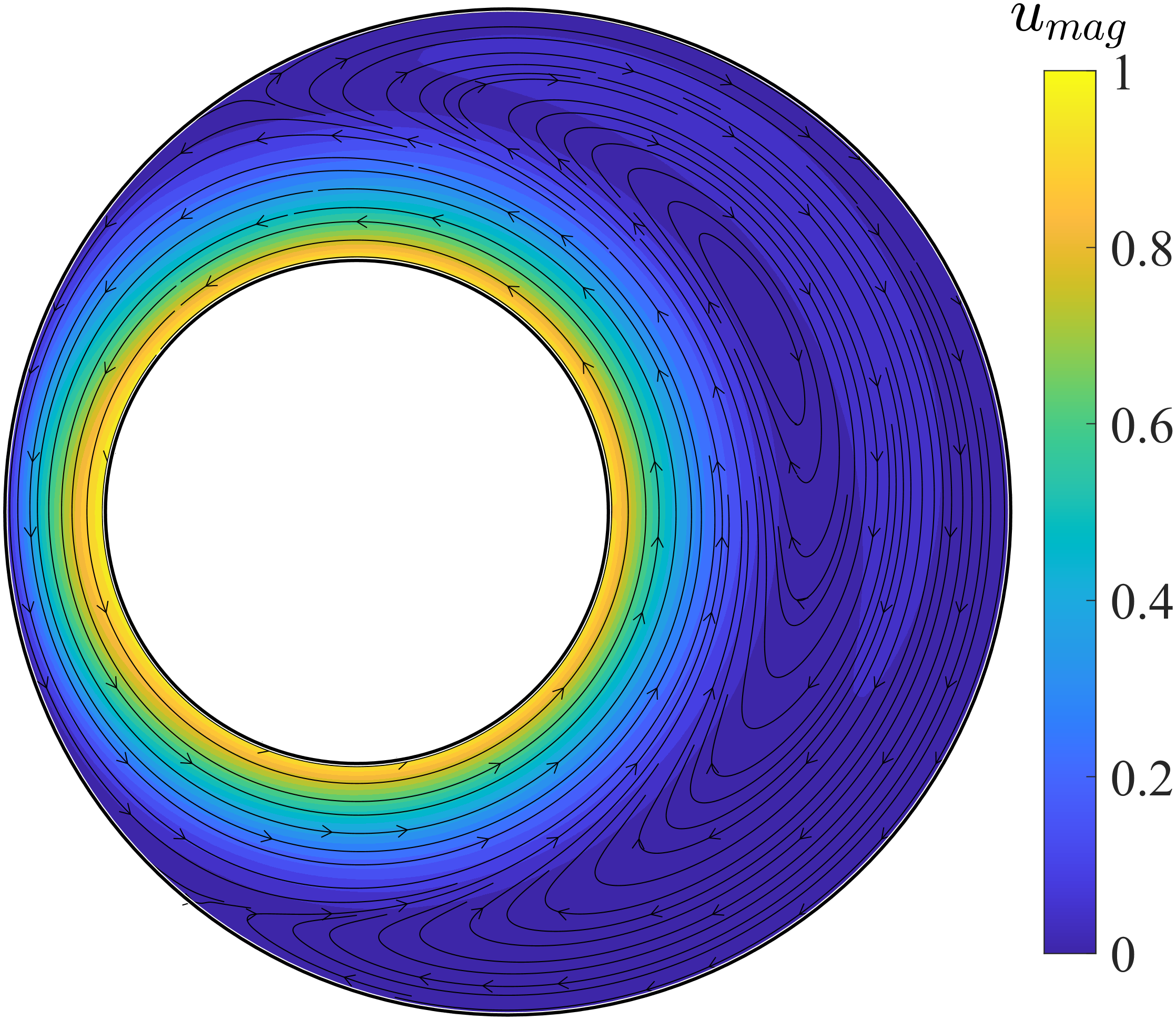}
        \caption{}
        \label{fig:base flow_C}
    \end{subfigure}
    \caption{Contours of base flow velocity magnitude overlayed with streamlines for three cases: a) $e = 0$ (concentric case), b) $e=0.3$ and $Re = 74$, and c) $e=0.6$ and $Re = 107$}
    \label{fig:base flow}
\end{figure}

In this section, we first demonstrate the algorithm to predict the global stability of Couette flow in the gap between two circular cylinders placed concentrically or eccentrically. The stability of concentric Taylor-Couette flow has been extensively studied beginning with the work of \citet{taylor1923viii}. In this geometry, the base flow streamlines are parallel, and the streamlines are circular (\cref{fig:base flow_A}). The tangential velocity, which is the only velocity component, is a function only of the radius ($r$), given as
\begin{equation}
      U_{\theta} = \left(\frac{\Omega r_i^2}{r_o^2-r_i^2}\right) \left(-r + \frac{r_o^2}{r}\right)  
\end{equation} 
To demonstrate our method, we use a distribution of scattered points shown in \cref{fig:point_dist}. The scattered points are described by coordinate pairs ($x$, $y$) and the base flow is specified as Cartesian ($x$) and ($y$) velocities from the above expression for the tangential velocity. We have also computed the base flow by numerically solving the Navier-Stokes equations in Cartesian coordinates using a meshless fractional step procedure reported by \citet{shahane2021high}. The numerically computed velocities differed from the analytical solution within 1.0e-6 in the L1 norm using 1677 scattered points. The global stability calculations for this case have been performed using the analytical base flow velocities. 

\begin{table}
  \begin{center}
\def~{\hphantom{0}}
  \begin{tabular}{cccc}
      No. of points, $n$  & Growth rate, $\omega_i$   &   CPU time [s] & Relative error [\%]\\[3pt]\hline
       ~771   & 0.032686 & ~~4.38 & 8.28\\
       1677   & 0.035506 & ~25.25 & 0.37\\
       2921  & 0.035598 & ~70.12 & 0.11\\
       4523   & 0.035613 & 193.24 & 0.06\\
       6460 & 0.035609 & 550.81 & 0.07\\
  \end{tabular}
  \caption{Grid independency for $Re = 74.924$ and $\alpha = 3.161$. $n$ denotes the number of scattered points. CPU time mentioned is the time required on an Intel Xeon(R) processor (2.5GHz)to calculate the first 100 eigenvalues with the smallest absolute value and the corresponding eigenvectors. Relative errors are compared with the value reported by \citet{marcus1984simulation}, $\omega_i = 0.035637$.}
  \label{tab:validation}
  \end{center}
\end{table}

\begin{figure}
    \centering
    \begin{subfigure}[b]{0.4\textwidth}
        \includegraphics[height = 50mm]{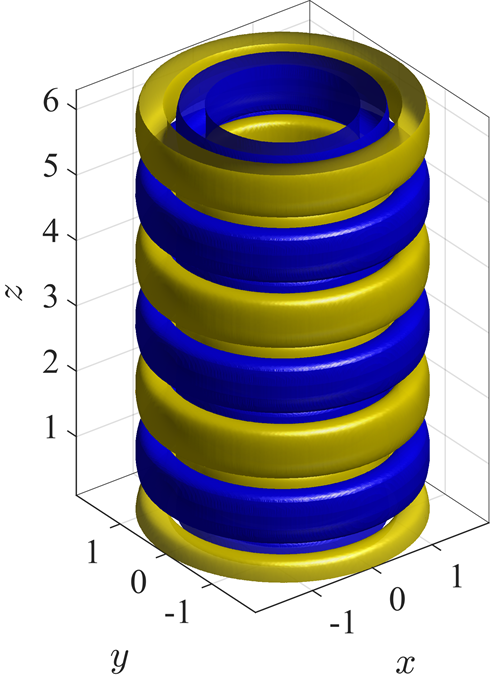}
        \caption{}
        \label{}
    \end{subfigure}
    \begin{subfigure}[b]{0.4\textwidth}
        \includegraphics[height = 45mm]{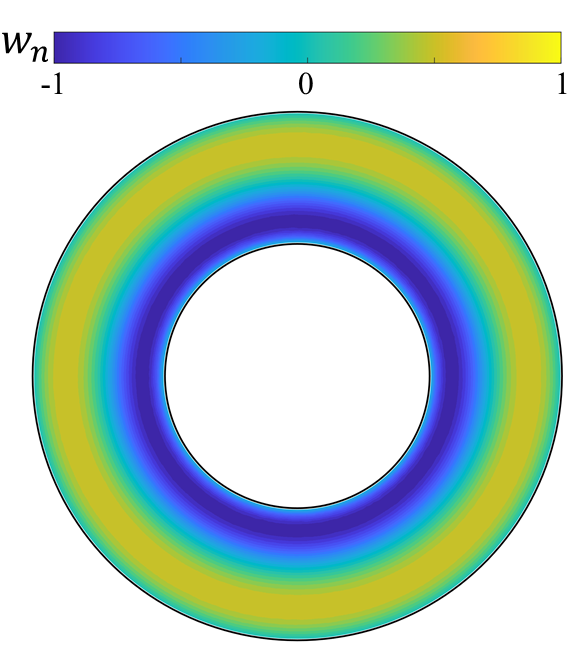}
        \caption{}
        \label{}
    \end{subfigure}
    \caption{Most unstable mode for the concentric case resulting in Taylor cells. a) Isosurfaces of normalised axial velocity $\Re(\hat{w_n}) = 0.2$ and $\Re(\hat{w_n}) = -0.2$, and b) contours of axial velocity, at $z = 0$, and $Re = 68.5$; Velocities are normalised with the maxima of $(\Re(\hat{w_n}))$. }
    \label{fig:e0_mode1}
\end{figure}

The $\boldsymbol{A}$ and $\boldsymbol{B}$ matrices in the global stability analysis use the Cartesian velocities and the physical coordinates of the scattered points. The effect of the number of scattered points is first investigated by computing the growth rate (the imaginary part of $\omega$) at $Re = 74.924$ and for a wave number $\alpha = 3.161$. This specific $Re$ and $\alpha$ are those computed by \citet{marcus1984simulation} using a pseudo-spectral algorithm. \Cref{tab:validation} shows our computed growth rate and relative error for different numbers of scattered points ($n$). For this eigenvalue, \citet{marcus1984simulation} gave a growth rate of 0.035637. Our results are within 0.4\% of this value with $n = 1677$, and within 0.11\% for $n = 2921$ scattered points. It can be seen that good accuracy is achieved with just 1677 points and requires only 25.5 seconds of CPU time of an Intel Xeon(R) processor (2.5GHz) to compute the first 100 eigenvalues and vectors. These times are significantly smaller than previously reported times for a typical global stability analysis. The first transition is predicted to be around $Re = 68.16$ (close to values reported by other researchers (\citet{fasel1984numerical,oikawa1989,leclercq2013temporal})). \Cref{fig:e0_mode1} shows the most unstable eigenvector for a supercritical $Re$ of 68.5, which is slightly past the $Re$ for transition. Since our stability analysis is global, we do not need to prescribe the azimuthal wave number. At this $Re$ and $\alpha$, we observe that the azimuthal wave number is zero for the most unstable mode. \Cref{fig:e0_mode1} shows the contours and iso-surfaces of the normalized axial eigenvector corresponding to the unstable mode, which illustrates the formation of the Taylor cells.

\begin{figure}
    \centering
    \begin{subfigure}[b]{\textwidth}
    \centering
        \includegraphics[width = 0.45\textwidth]{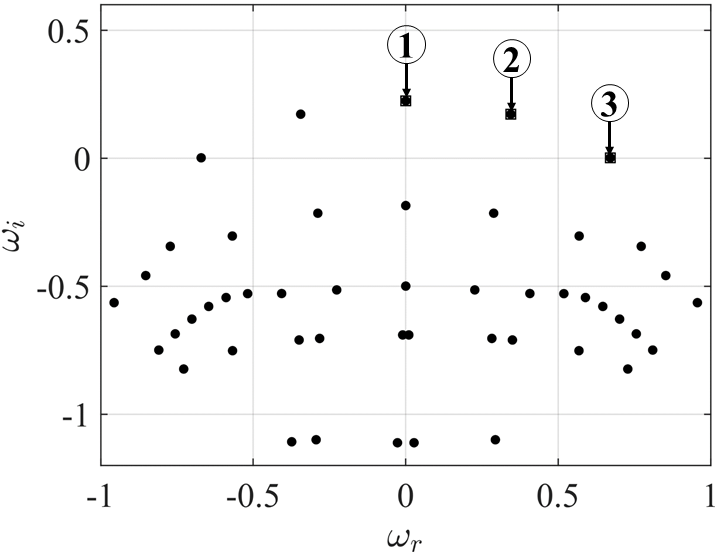}
        \caption{}
        \label{fig:e0_re110_modes_a}
    \end{subfigure}
    \par \bigskip
    \begin{subfigure}[b]{\textwidth}
    \centering
        \includegraphics[height = 45mm]{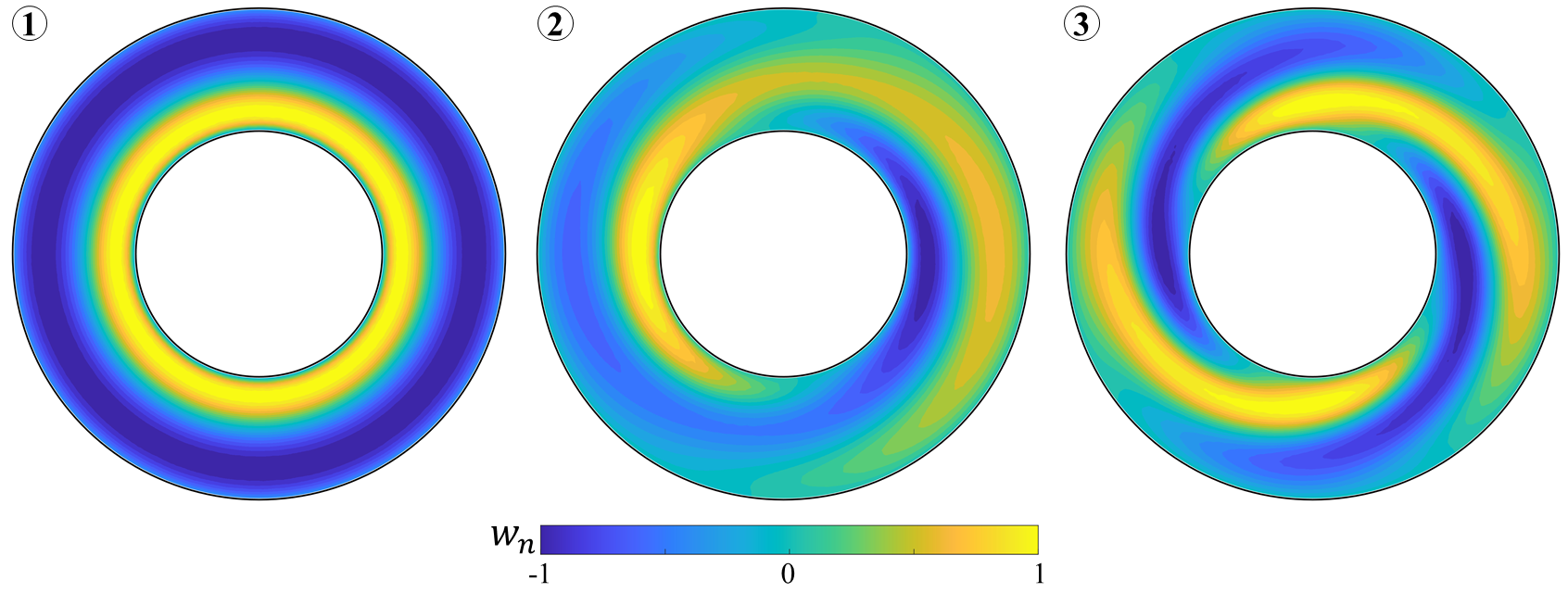}
       \caption{}
    \end{subfigure}
    \begin{subfigure}[b]{\textwidth}
    \centering
        \includegraphics[height = 50mm]{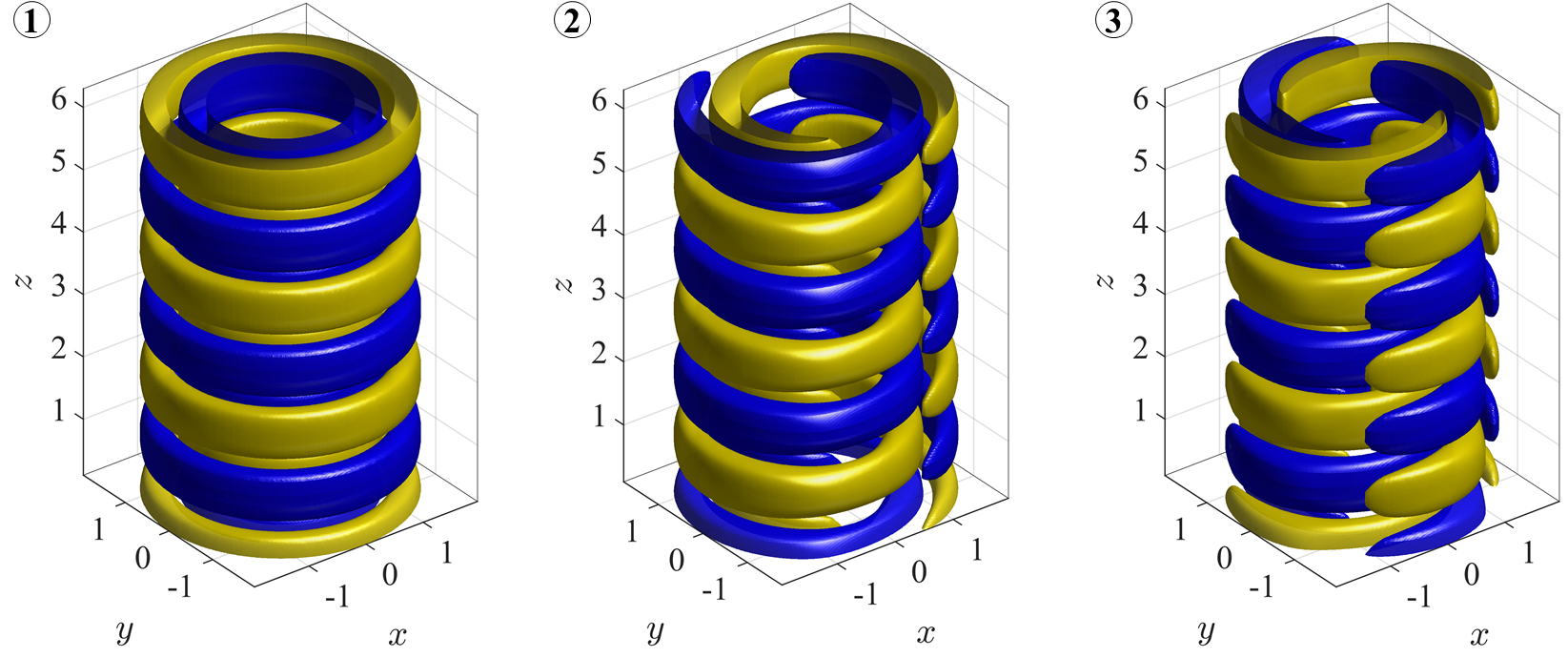}
        \caption{}
        \label{fig:e0_re110_modes_b}
    \end{subfigure}
    \caption{a) Eigenvalue spectrum with the three unstable modes marked (two of the unstable modes have a complex conjugate) for a $Re = 150$ and $\alpha = 3.15$, b) contours of normalized axial velocity at $z=0$ corresponding to the three marked modes,  and c) isosurfaces of normalized axial velocity for the values of 0.2 and -0.2 corresponding to the marked modes.}
    \label{fig:e0_re110_modes}
\end{figure}

\begin{figure}
    \centering
        \begin{subfigure}[b]{0.8\textwidth}
        \includegraphics[width = \textwidth]{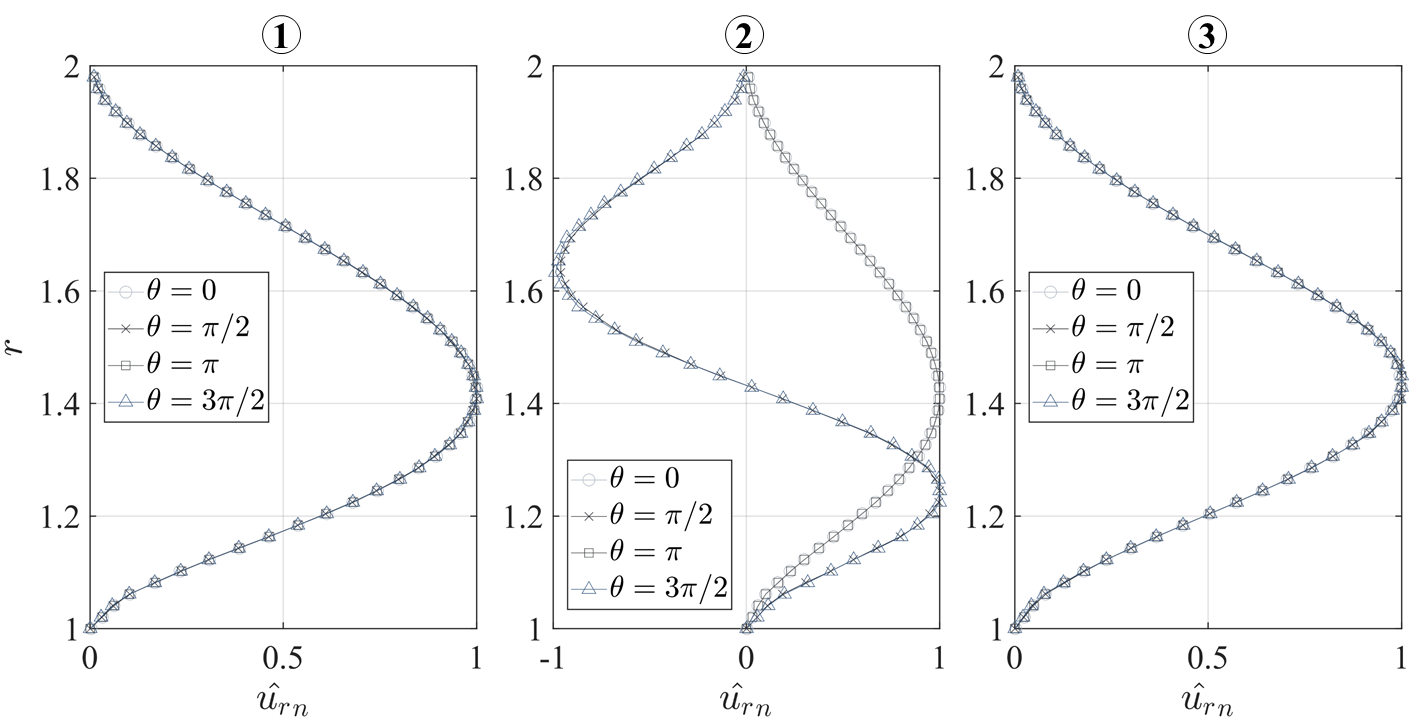}
        \caption{}
        \label{}
    \end{subfigure}
    \par \bigskip
    \begin{subfigure}[b]{0.8\textwidth}
        \includegraphics[width=\textwidth]{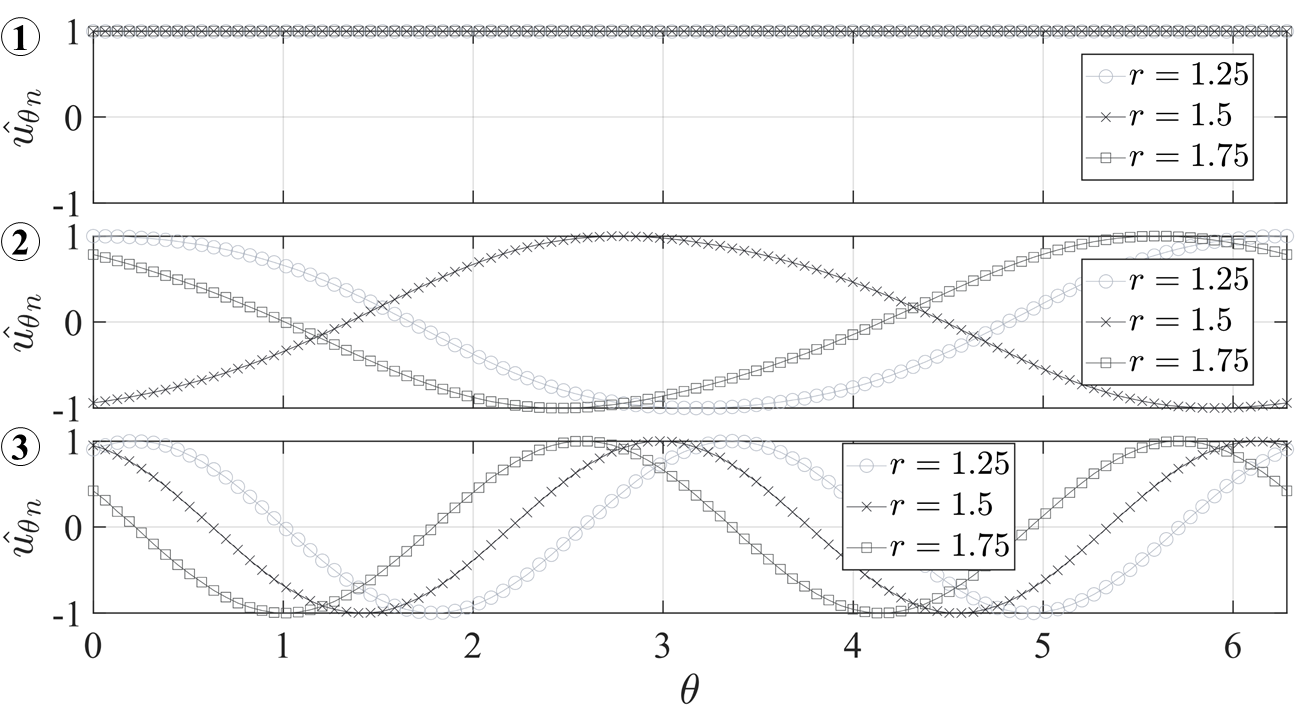}
        \caption{}
    \end{subfigure}
    \caption{Normalised radial and tangential velocity profiles at different radial and angular locations. a) Radial velocity profiles normalized with the maxima of the profile at angular locations $\theta =0$, $\theta =\pi/2$, $\theta = \pi$, and $\theta = 3\pi/2$ for the modes [from left] 1, 2, and 3. b) Tangential velocity profiles normalized with the maxima of the profile at radial locations $r = 1.25$, $r=1.5$, and $r=1.75$  for the modes [from top] 1, 2, and 3. }
    \label{fig:norm_velocity_TC}
\end{figure}

We next performed global stability at a higher $Re$ of 150 and for $\alpha = 3.15$. In all the previous local stability analyses of Taylor-Couette flow, the normal mode form was represented in cylindrical coordinates as $\psi(r)\exp{\iota(\alpha z + \beta\theta - \omega t)}$, where $\alpha$ represents the wavenumber in the $z$ direction, $\beta$ is an integer representing the number of waves present along the $\theta$ direction.  In the present analysis, we do not prescribe the number of waves, but rather allow a general functional representation in $x$ and $y$ (Please refer to the normal mode form given in equation \ref{eqn:normal_mode}). The eigenspectrum computed for a $Re$ of 150 and $\alpha=3.15$ and the corresponding eigenfunctions (of the three selected modes) are given in \cref{fig:e0_re110_modes}. We are particularly interested in the first few modes having maximum growth rate. The first three modes are selected. Interestingly, the first is a two-dimensional mode as the perturbation is axisymmetric. The second mode is non-axisymmetric with one wave encircling the inner cylinder. This corresponds to $\beta=1$ of the normal mode form in the cylindrical coordinate system.  The third mode is again non-axisymmetric with two waves encircling the inner cylinder, where $\beta=2$. In \cref{fig:norm_velocity_TC}, we have plotted the eigenfunctions of three modes at different pre-selected radial and angular locations. These are radial and tangential velocities normalized with the maxima of the profile. The number of waves encircling the inner cylinder is obvious from the plots. The merit of the current global analysis method is that the azimuthal modes directly come out of the analysis without pre-assignment. All modes can be directly obtained from a single analysis. Lastly, we compared results for a different geometry given in \cite{oikawa1989stability}. For a radius ratio, $\eta = 0.83$ and $\alpha = 15.415$, we obtain a critical Reynolds number of $532$ with the growth rate, $\omega_i < 1e-3$. \citet{oikawa1989} (fig. 4 of their paper) give $Re_c = 109.1$ for $\eta = 0.829$, and $\alpha_c = 3.16$, which is equivalent to our computed value of 532 because of the difference in length scales used in the definition of $Re$).

Although the concentric Couette flow can be studied by local stability analysis, its variation with the inner cylinder placed eccentrically is not amenable to such a method because it does not have parallel streamlines. Several previous studies have performed local stability analyses of the eccentric configuration by mapping the domain to a concentric configuration. A summary of these works is given in \cref{sec:intro}. As before, the base flow for this geometry is computed at the scattered points and is described in terms of the two Cartesian velocities along the $x$ and $y$ directions. The base flow here was computed using a semi-implicit meshless solver, developed by \cite{memphys}. We append a polynomial of degree 5 to the RBF and march in time to a steady state with a convergence tolerance of 1e-10. These base flow results at the same set of scattered points are used also for stability analysis. The computation time for base flow for an eccentric case at a Reynolds number, $Re = 107$, eccentricity, $e = 0.6$, with the appended polynomial of degree $5$, was 523 s for a point cloud of $4173$ points to reach steady-state tolerance of $1e-10$ on a single core of Intel Xeon(R) Gold 6248 CPU clocked at 2.50GHz. \Cref{fig:base flow_B,fig:base flow_C} show the streamlines computed for two eccentricities ($e = 0.3$ and $e = 0.6$). 

Eccentric Taylor-Couette flow was previously analyzed in detail by \citet{oikawa1989stability}, \citet{leclercq2013temporal}, and their results are considered for comparison in the present work. It is to be noted that the eccentric placement of the inner cylinder creates non-parallel flow within the annulus. The higher eccentricity causes the flow to generate a separation bubble in the wide gap region which can be seen in \cref{fig:base flow_C}. The adverse and favorable pressure gradients also affect the stability of the flow. Since the classical linear stability theory assumes the base flow to be parallel, a global stability analysis is mandated for this configuration.

\begin{figure} 
    \centering
    \begin{subfigure}[b]{0.45 \textwidth}
    \centering
        \includegraphics[width = \textwidth]{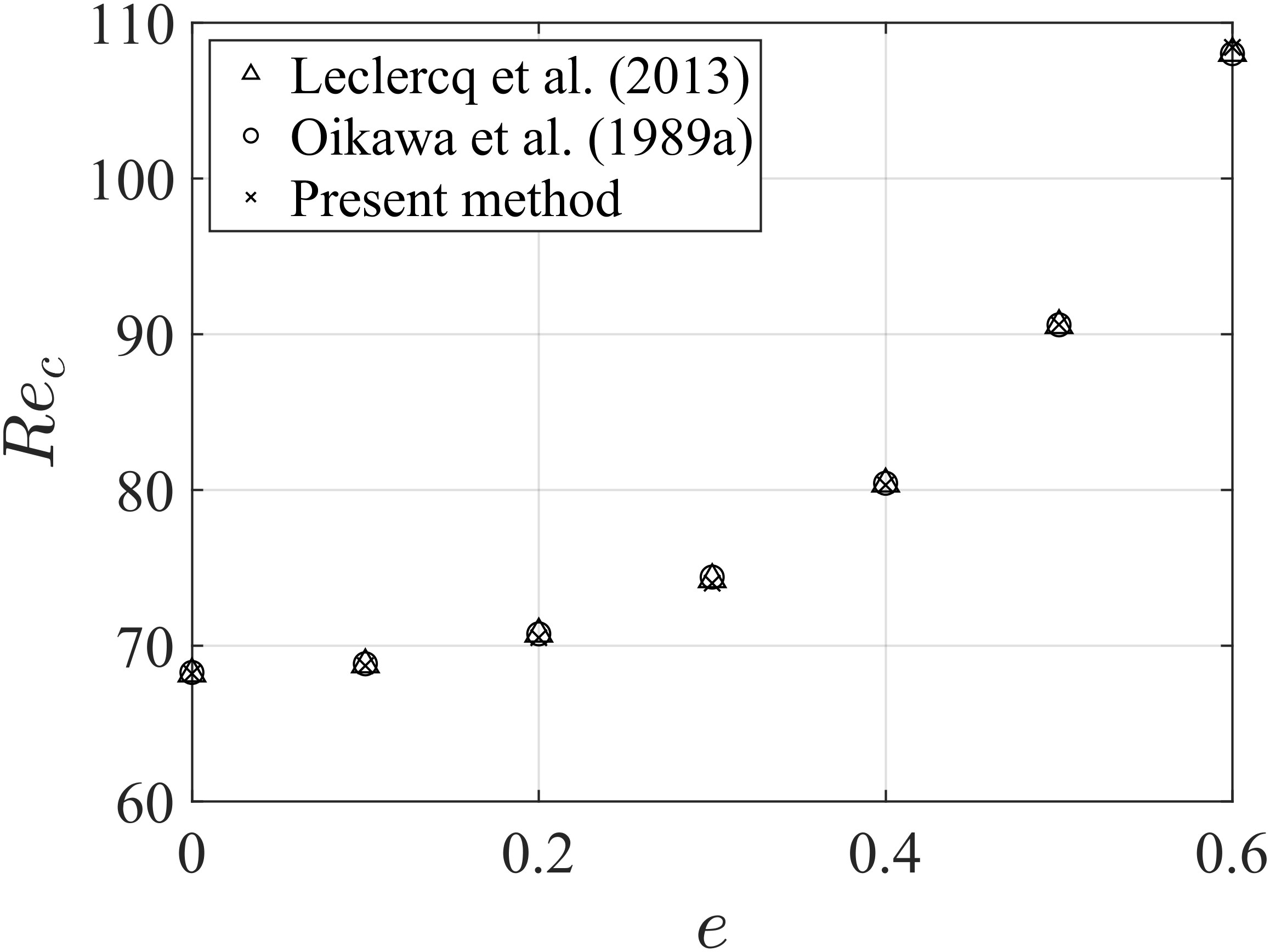}
        \caption{}
        \label{fig:eccentric_Re_alpha_comparison_a}
    \end{subfigure}
    \hfill
    \begin{subfigure}[b]{0.45 \textwidth}
    \centering
        \includegraphics[width = \textwidth]{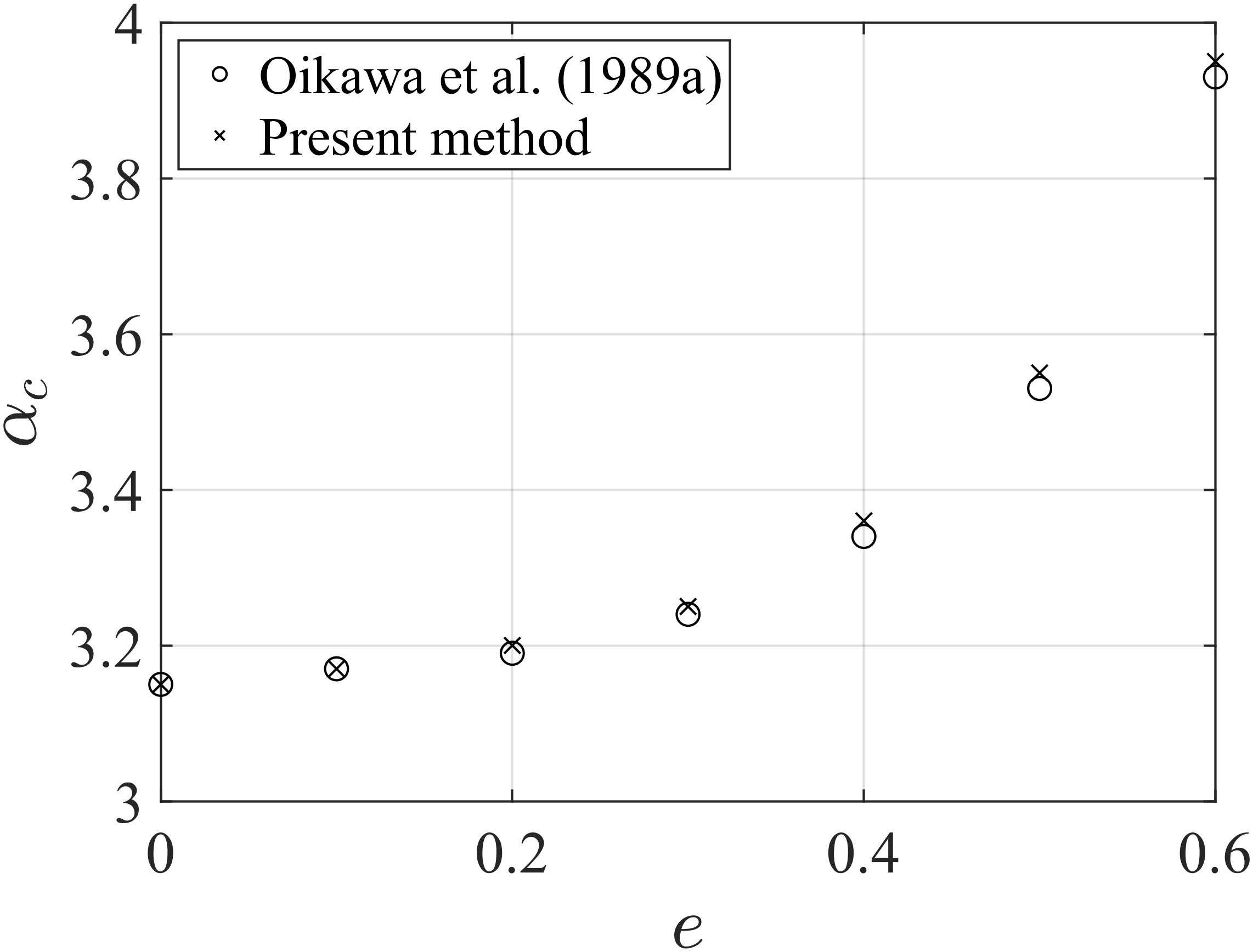}
        \caption{}
        \label{fig:eccentric_Re_alpha_comparison_b}
    \end{subfigure}
    \caption{a) Critical Reynolds numbers ($Re_c$) at various eccentricities ($e$) compared with \citet{oikawa1989} and \cite{leclercq2013temporal}, b) Critical wavenumbers ($\alpha_c$) at various eccentricities compared with \citet{oikawa1989}}
    \label{fig:eccentric_Re_alpha_comparison}
\end{figure}

\begin{figure}
    \centering
    \begin{subfigure}[b]{0.45 \textwidth}
    \centering
        \includegraphics[width = \textwidth]{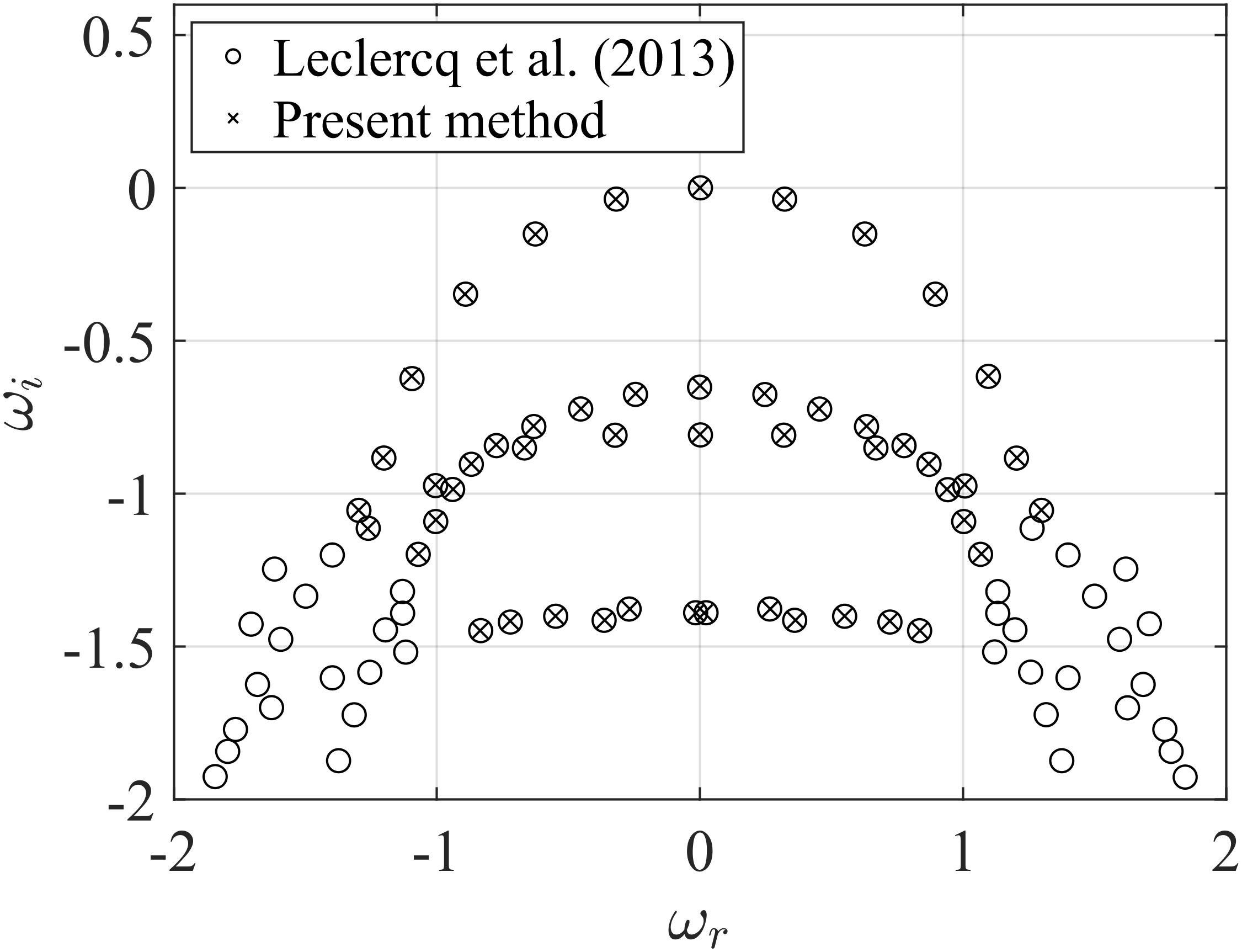}
        \caption{}
        \label{}
    \end{subfigure}
    \hfill
    \begin{subfigure}[b]{0.45 \textwidth}
    \centering
        \includegraphics[width = \textwidth]{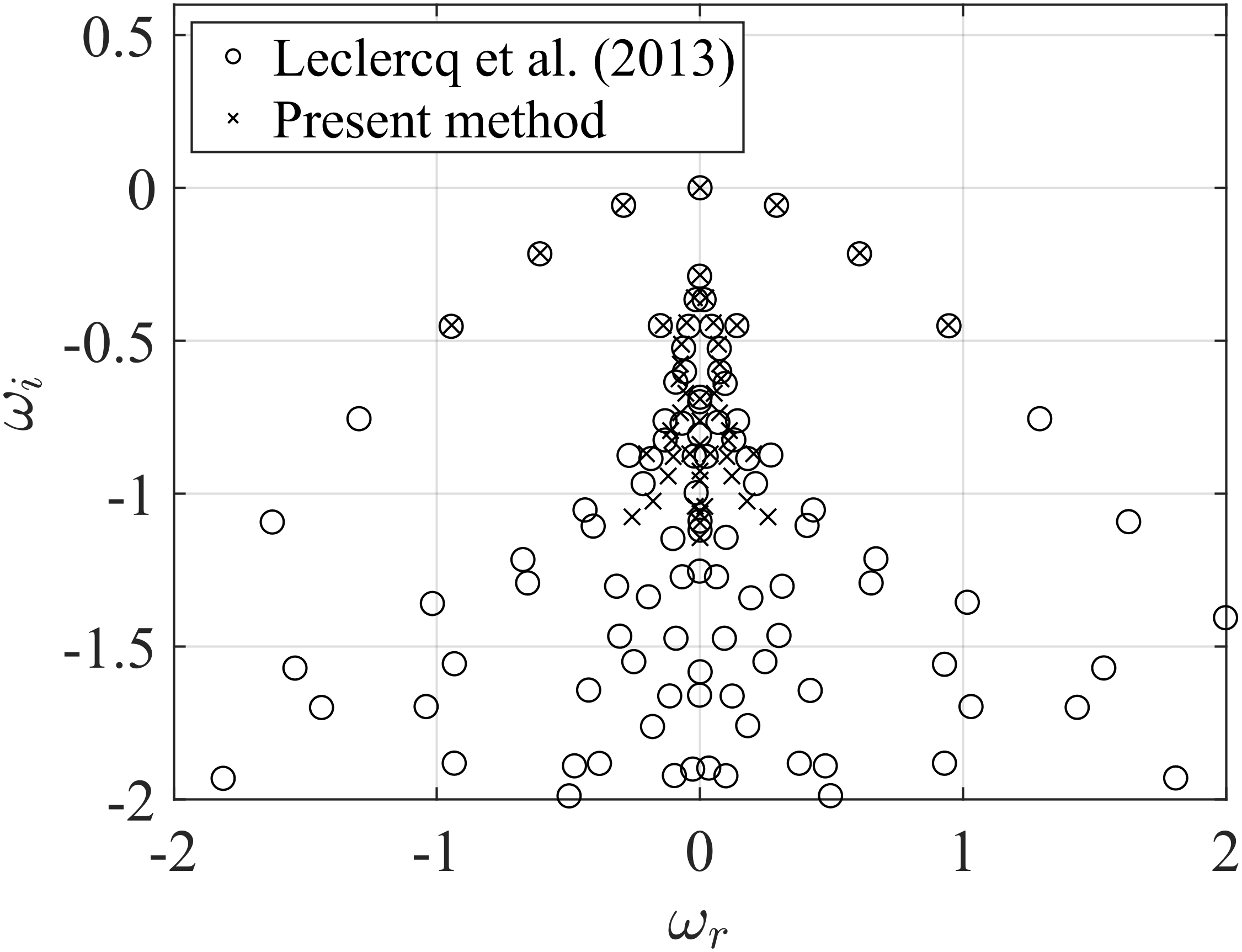}
        \caption{}
        \label{}
    \end{subfigure}
    \caption{Eigenvalue spectrum compared with the data extracted from figures in \cite{leclercq2013temporal} for a) $e=0$ and b) $e=0.5$ at corresponding critical Reynolds numbers }
    \label{fig:spectrum_comparison}
\end{figure}

\Cref{fig:eccentric_Re_alpha_comparison_a} provides a plot of the critical Reynolds numbers computed for different eccentricities and compares them with results given by \citet{leclercq2013temporal} and \citet{oikawa1989}. \Cref{fig:eccentric_Re_alpha_comparison_b} plots the axial wavenumber against the eccentricity ratio. Our calculations agree very well with their results at all values of eccentricities. \Cref{fig:spectrum_comparison} compares the first 50 eigenmodes of the concentric case and an eccentric case ($e=0.5$) with \citet{leclercq2013temporal}. 
The most critical modes with close to zero growth rate are coincident with computations of \citet{leclercq2013temporal}.


\begin{figure}
    \centering
    \begin{subfigure}[b]{0.25 \textwidth}
        \includegraphics[width = \textwidth]{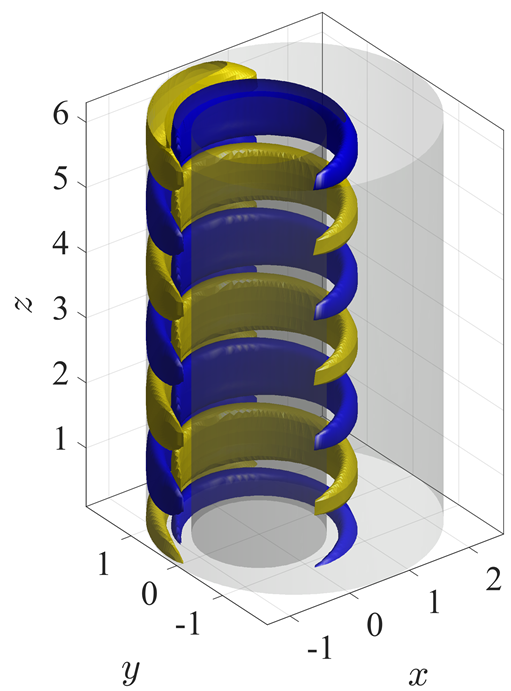}
        \caption{}
        \label{fig:e60_mode1_a}
    \end{subfigure}
    \begin{subfigure}[b]{0.35 \textwidth}
        \includegraphics[width = \textwidth]{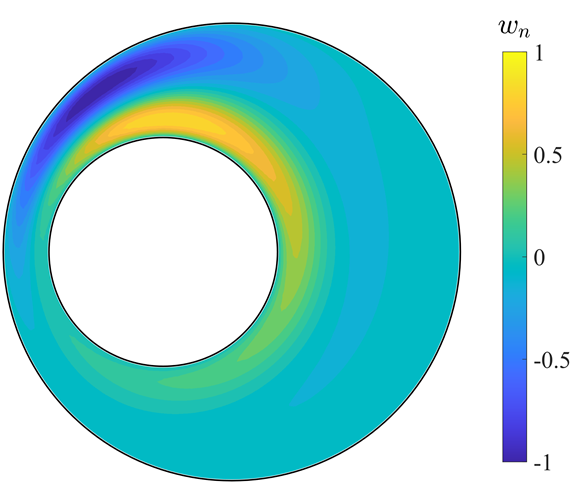}
        \caption{}
        \label{fig:e60_mode1_b}
    \end{subfigure}
    \begin{subfigure}[b]{0.35 \textwidth}
        \includegraphics[width = \textwidth]{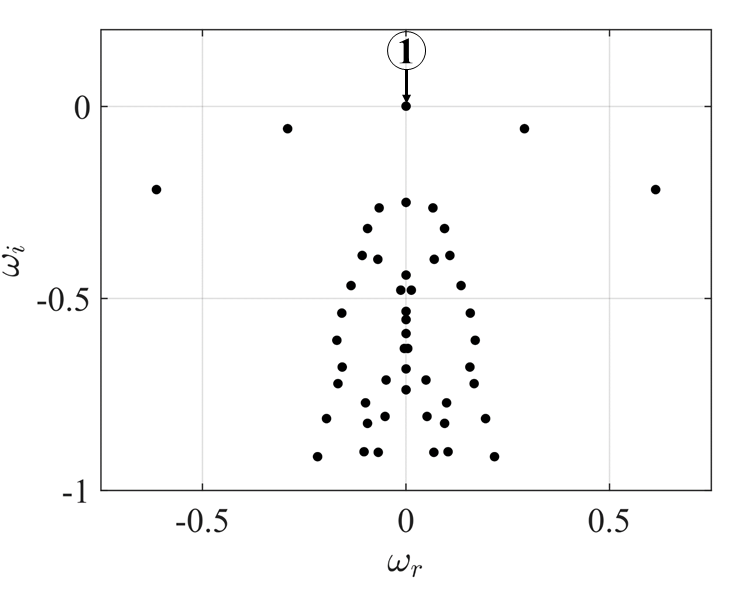}
        \caption{}
        \label{fig:e60_mode1_c}
    \end{subfigure}
    \caption{ Most unstable mode for an eccentricity of $e=0.6$, $Re = 108.5$, and $\alpha = 3.95$. a) Three-dimensional structure of the mode plotted as an isosurface of axial velocity $\Re(w_n) = 0.2$ and $\Re(w_n) = -0.2$; b) contour plot of the axial velocity at $z=0$ plane; c) eigen spectrum of the complex frequencies with frequency on the x-axis and growth rate plotted on the y-axis.}
    \label{fig:e60_mode1_and_spectrum}
\end{figure}
\begin{figure}
    \centering
     \begin{subfigure}[b]{0.4 \textwidth}
        \centering
        \includegraphics[height = 50mm]{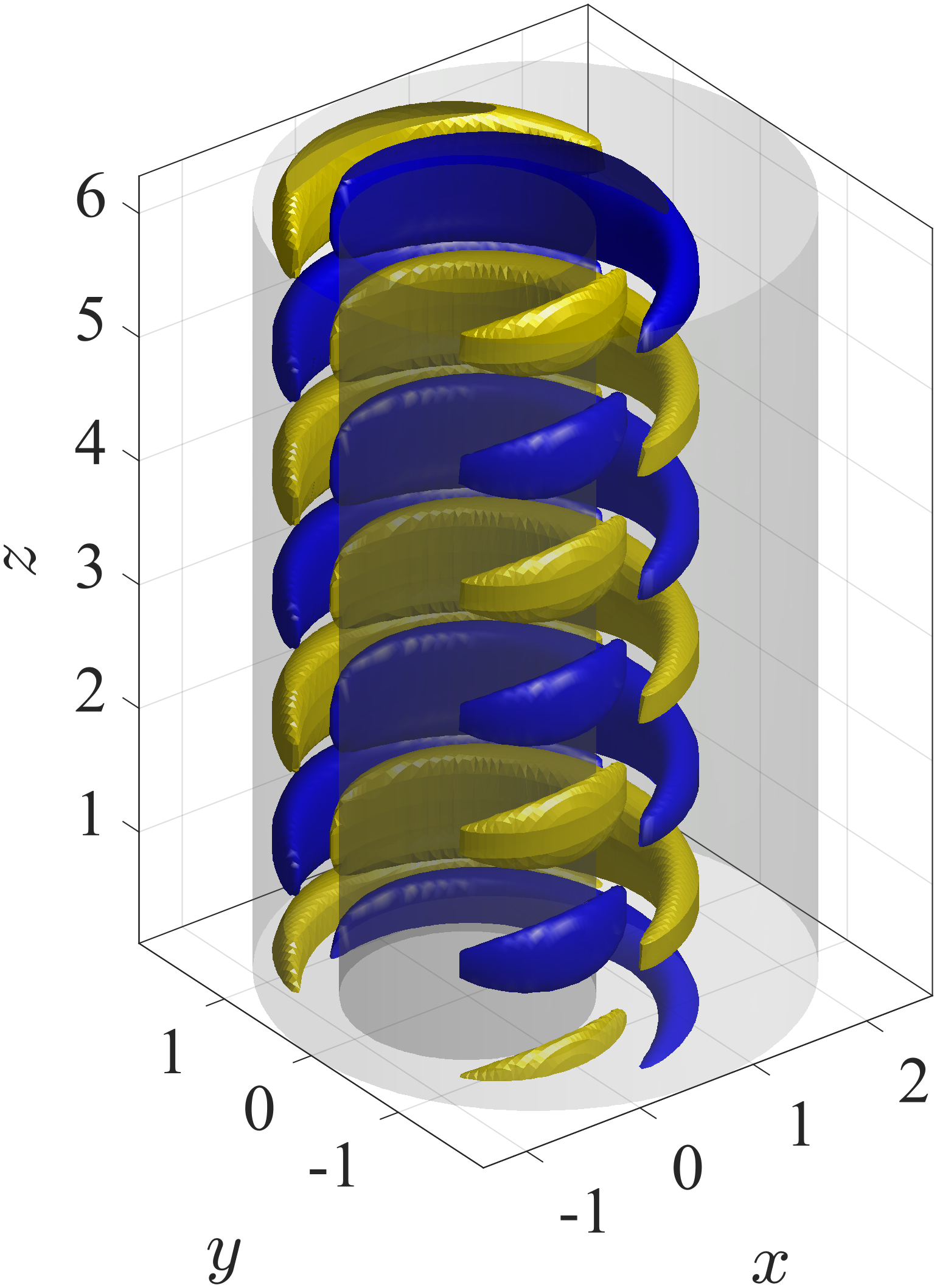}
        \caption{}
        \label{fig:e60_mode2_isosurface}
    \end{subfigure}
    \begin{subfigure}[b]{0.4 \textwidth}
    \centering
        \includegraphics[height= 42mm]{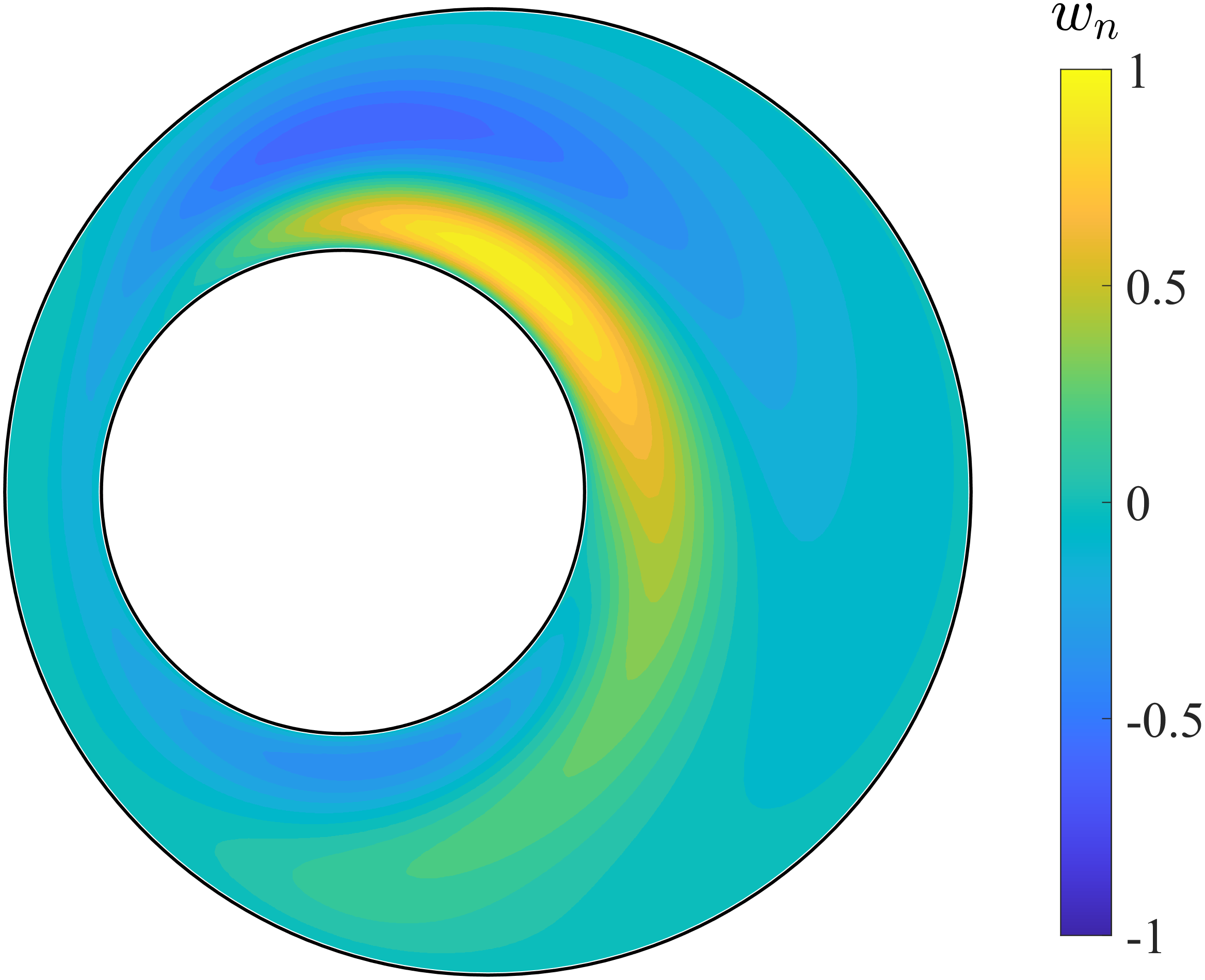}
        \caption{}
        \label{fig:e60_mode2_contour}
    \end{subfigure}
   \caption{Second mode (stable) for the eccentric case with $e=0.6$, $Re = 108.5$ and $\alpha = 3.95$. a) Isosurfaces of normalized axial velocity $\Re(w_n) = 0.2$ and $\Re(w_n) = -0.2$. b) Contour plot of $\Re(w_n)$ at $z = 0$.}
    \label{fig:e60_mode2}
\end{figure}

\Cref{fig:e60_mode1_a} presents the most unstable mode for an eccentric case of $e=0.6$, $Re = 108.5$, and $\alpha = 3.95$. The base flow being non-parallel, a local stability analysis with prescribed periodic wave numbers in $\theta$ direction, would not give us the correct modes. A global stability analysis removes this requirement of prescribing the angular modes and retrieves complete information of the disturbance field in the plane even if it does not have a wavelike nature. We observe that for eccentric placement of the inner cylinder, there need not be only one constant wave mode around the cylinder. \Cref{fig:e60_mode2} presents the contours and isosurfaces of normalized axial velocity of the second mode (which is seen to be stable) of the same case. It can be observed that the wave nature of the perturbed field is similar to the $\beta=1$ mode for the concentric case, however, it is distorted because of the effects of eccentricity.

\section{Stability of Couette flow in an elliptical enclosure}
\label{sec:results}


We now consider a new geometry of Taylor-Couette flow for which only limited information exists. In this configuration, the outer enclosure is an ellipse, thus forming a different shape of the gap region. We have recently studied this flow with the meshless method (\citet{unnikrishnan2022shear,unnikrishnan2024taylor}) varying the aspect ratio of the ellipse and the eccentricity of placement of the inner cylinder. The observed flow patterns are described in \cite{unnikrishnan2022shear}. Using a procedure similar to that used in \cref{sec:validation}, we determine the critical modes and critical axial wavenumber by searching in ($Re,\alpha$) space. To our knowledge, this is the first study to determine the stability characteristics of Couette flow with an elliptical outer enclosure. 

\begin{figure}
    \centering
        \begin{subfigure}[b]{\textwidth}
        \includegraphics[width = \textwidth]{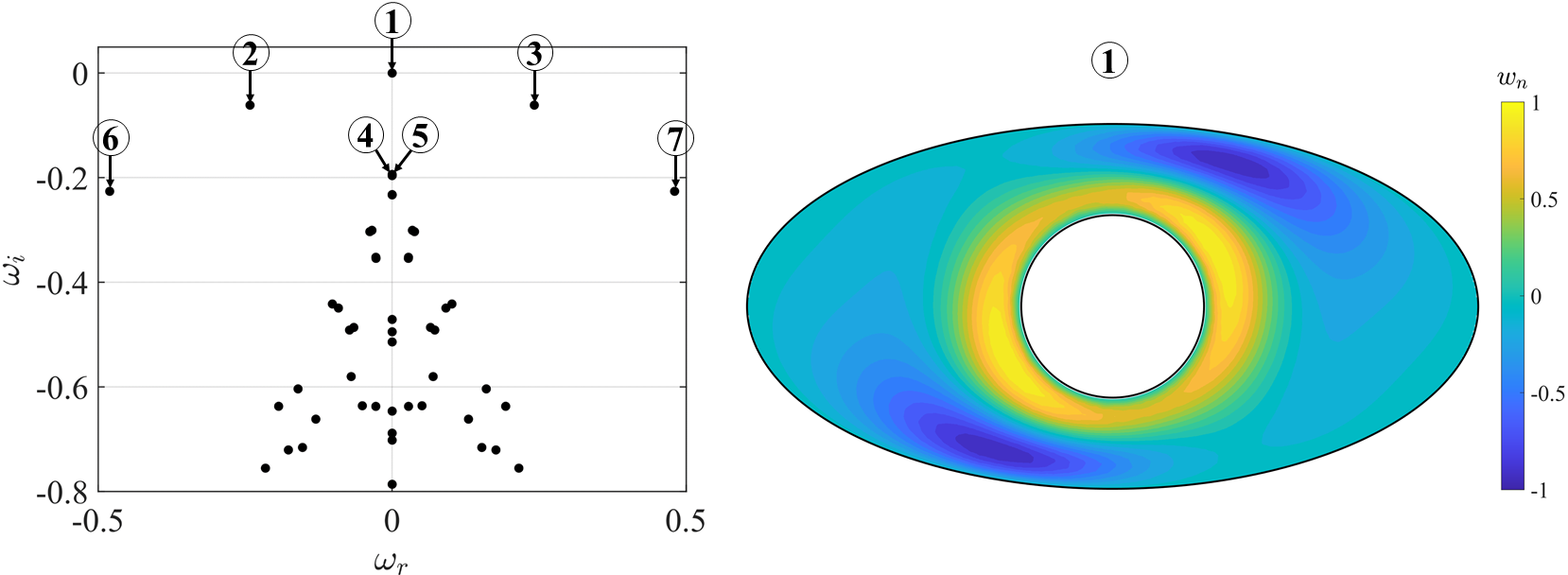}
        \caption{}
        \label{}
    \end{subfigure}
    \par \bigskip
    \begin{subfigure}[b]{0.95\textwidth}
     \includegraphics[width=\textwidth]{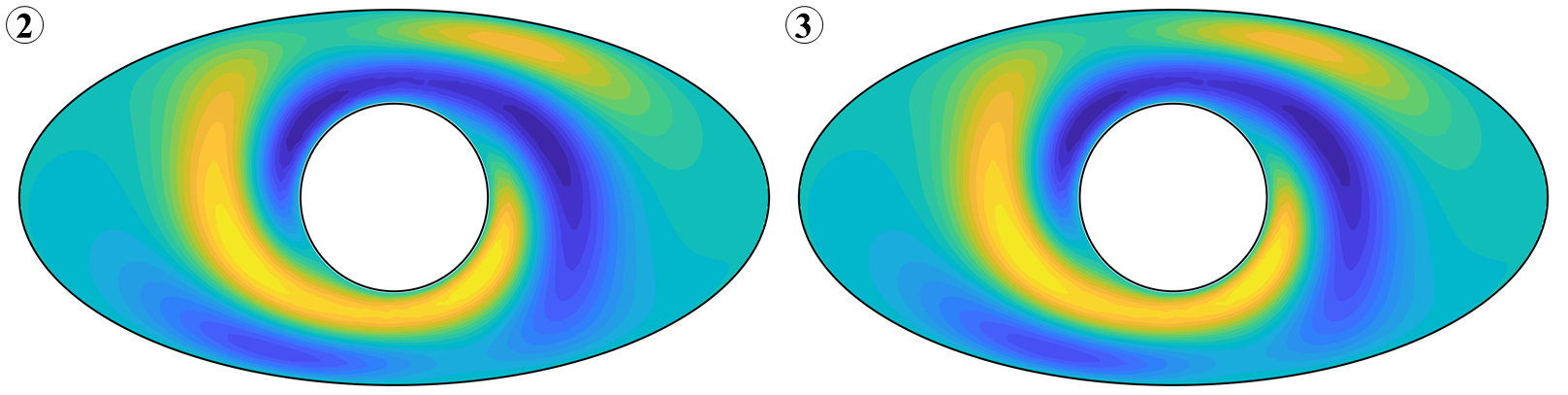}
    \end{subfigure}
    \begin{subfigure}[b]{0.95\textwidth}
     \includegraphics[width=\textwidth]{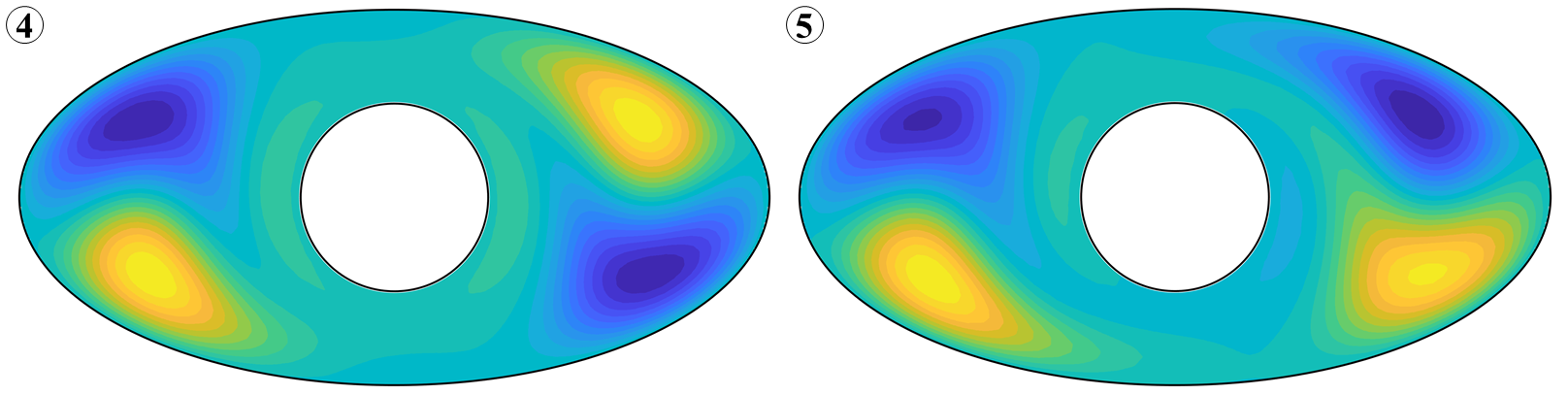}
    \end{subfigure}
    \begin{subfigure}[b]{0.95\textwidth}
     \includegraphics[width=\textwidth]{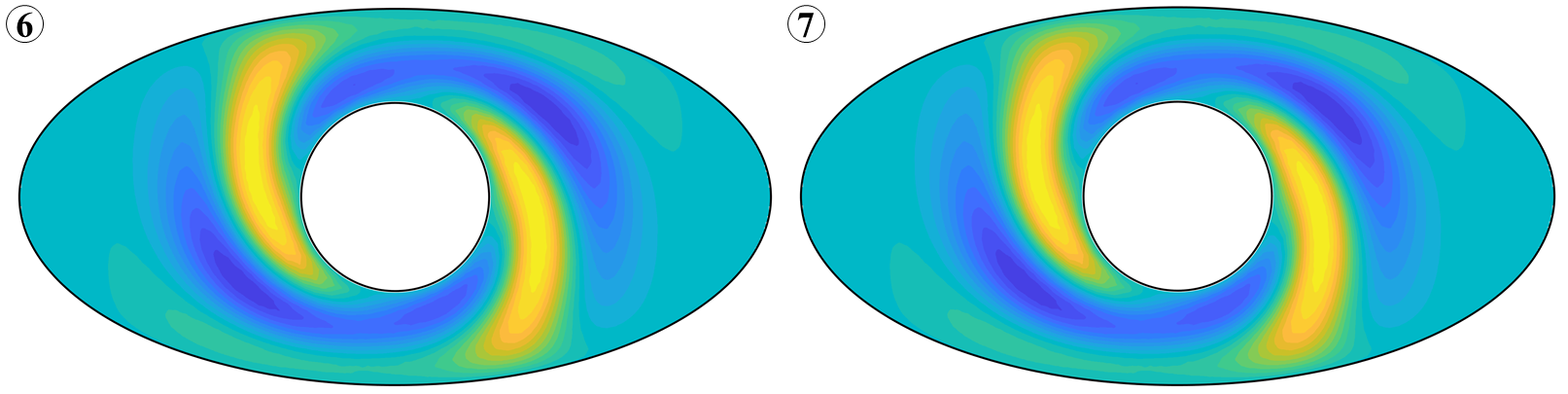}
     \caption{}
    \end{subfigure}
    \caption{a) The eigenvalue spectrum and the corresponding axial velocity contours at the critical Reynolds number 48.35 and the critical wave number $\alpha = 2.088$, where the growth rate of critical mode (marked 1 in the figure) is -1e-5. b) Representation of stable modes through normalized axial velocity contours at $z=0$ }
    \label{fig:elliptic_modes}
\end{figure}

\begin{figure}
    \centering
        \begin{subfigure}[b]{0.8\textwidth}
        \includegraphics[width = \textwidth]{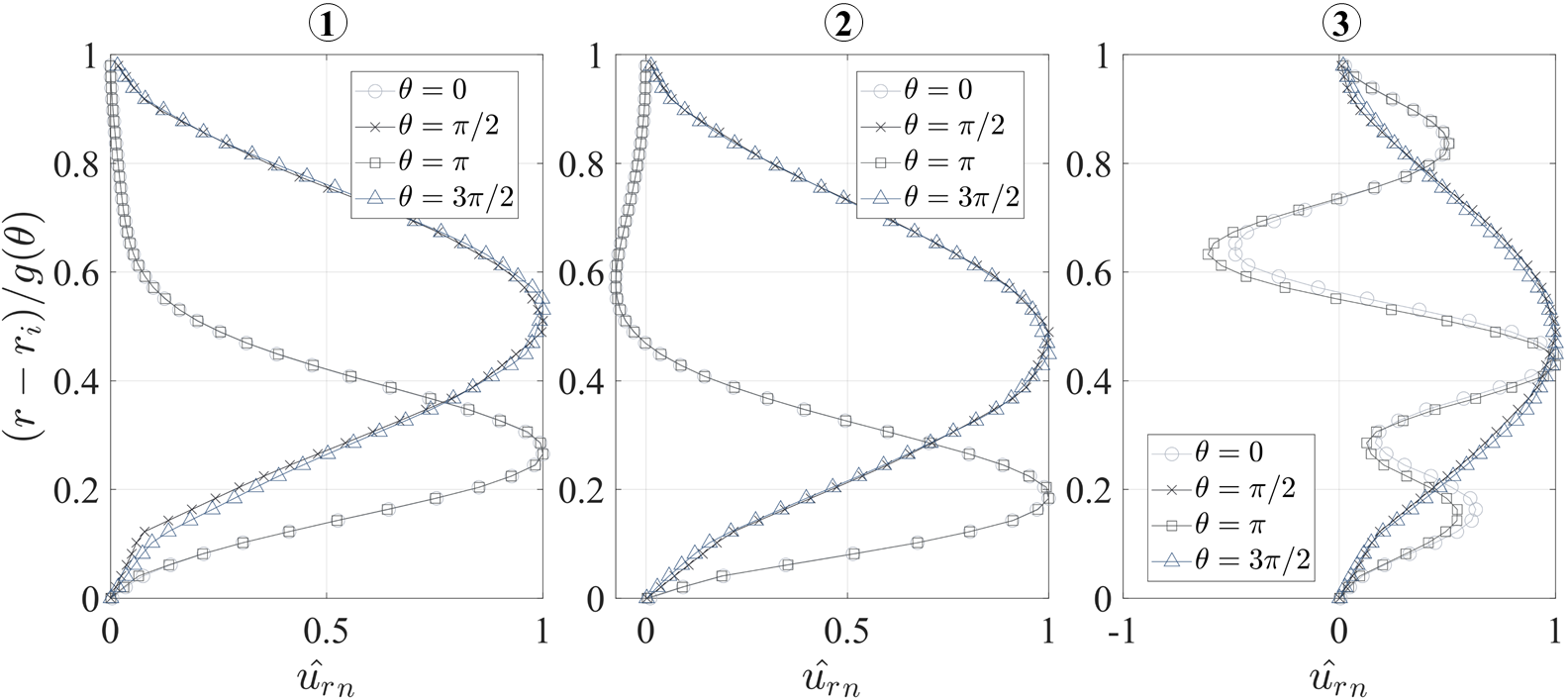}
        \caption{}
        \label{fig:ur_ut_elliptic_modes_a}
    \end{subfigure}
    \par \bigskip
    \begin{subfigure}[b]{0.8\textwidth}
        \includegraphics[width=\textwidth]{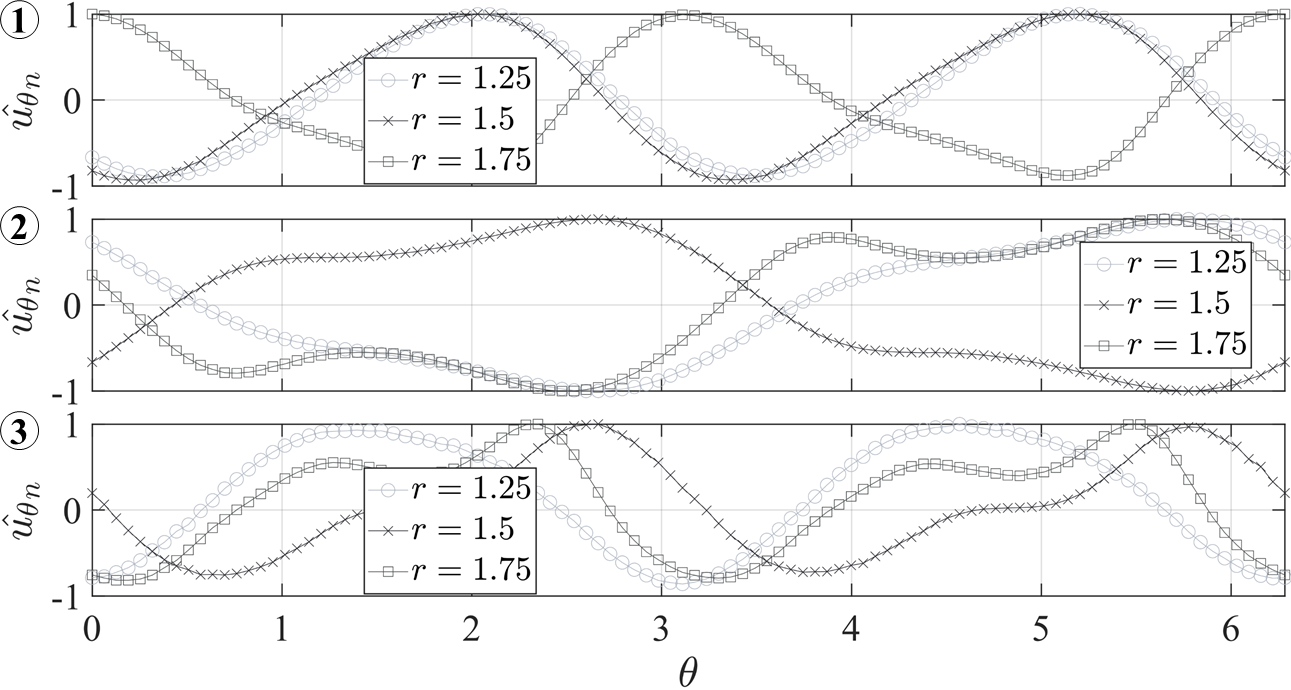}
        \caption{}
        \label{fig:ur_ut_elliptic_modes_b}
    \end{subfigure}
    \caption{a) Normalised radial velocity profiles of the first 3 modes at [from left] $\theta =$ 0, $\pi/2$, $\pi$, and $3\pi/2$; and b) normalised tangential velocity profiles at [from top] $r =$ 1.25, 1.5, and 1.75 at $Re_c = 48.35$ and $\alpha_c = 2.088$. $g(\theta)$ in a) represents the gap between the cylinders at the corresponding angular location.}
    \label{fig:ur_ut_elliptic_modes}
\end{figure}

\begin{table}
\begin{center}
\def~{\hphantom{0}}
\begin{tabular}{ccc}
\hspace{1cm}\textbf{e}\hspace{1cm} & \hspace{1cm}\textbf{$Re_c$}\hspace{1cm} & \hspace{1cm}\textbf{$\alpha_c$}\hspace{1cm} \\ [3pt]\hline
0.0~ & 48.35 & 2.088 \\ 
0.25 & 48.63 & 2.095 \\ 
0.5~ & 49.21 & 2.11~ \\ 
0.75 & 50.84 & 2.20~ \\ 
1.0~ & 53.0~ & 2.30~ \\ 
1.25 & 56.17 & 2.42~ \\ 
\end{tabular}

\caption{Critical Reynolds number and axial wave number for Taylor-Couette flow in an elliptical enclosure with an aspect ratio of 2.}
\label{tab:critical_elliptic}
\end{center}
\end{table}
\begin{figure}
    \centering
        \begin{subfigure}[b]{0.45\textwidth}
        \includegraphics[width = \textwidth]{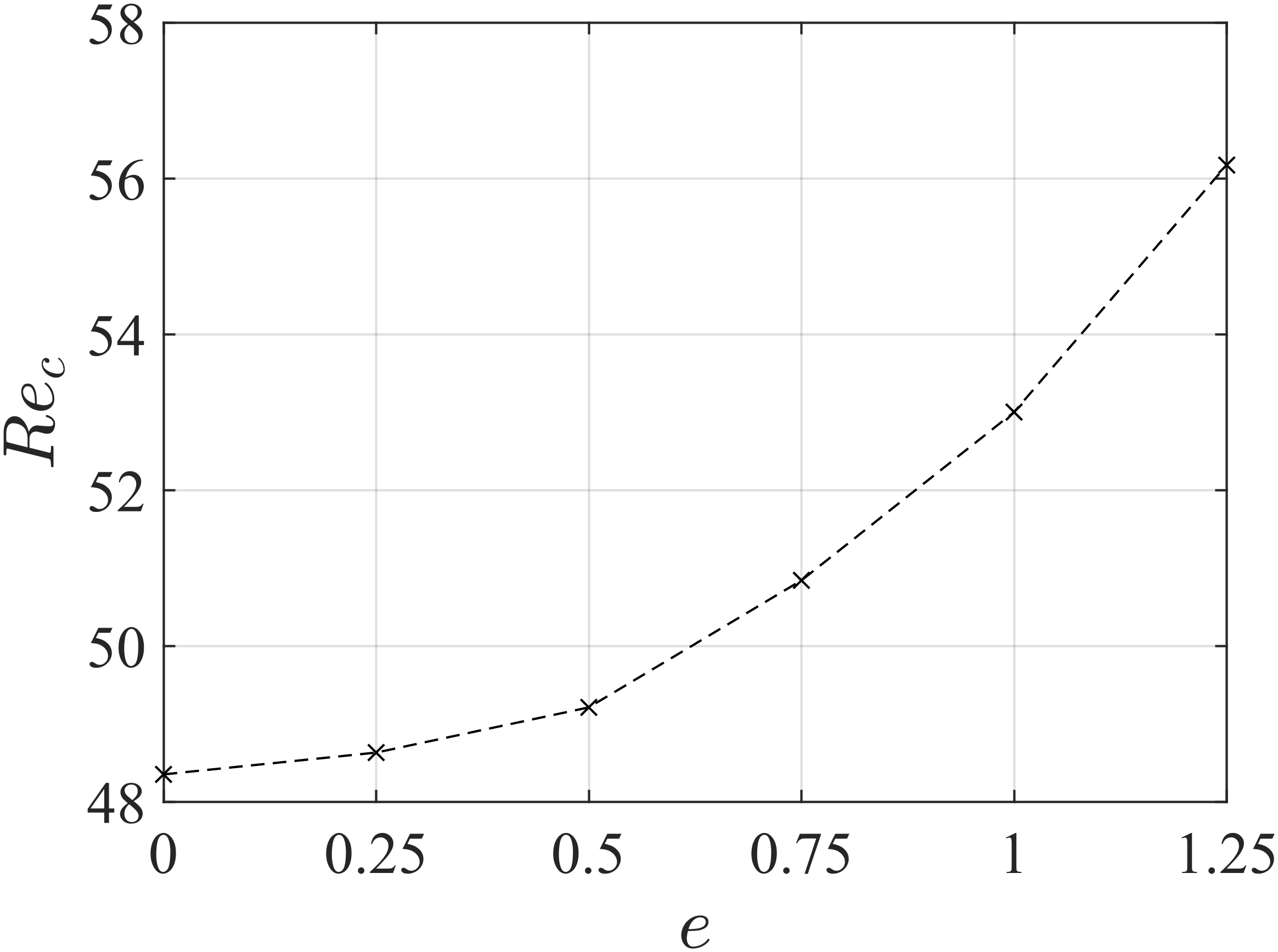}
        \caption{}
        \label{}
    \end{subfigure}
    \hfill
    \begin{subfigure}[b]{0.45\textwidth}
        \includegraphics[width=\textwidth]{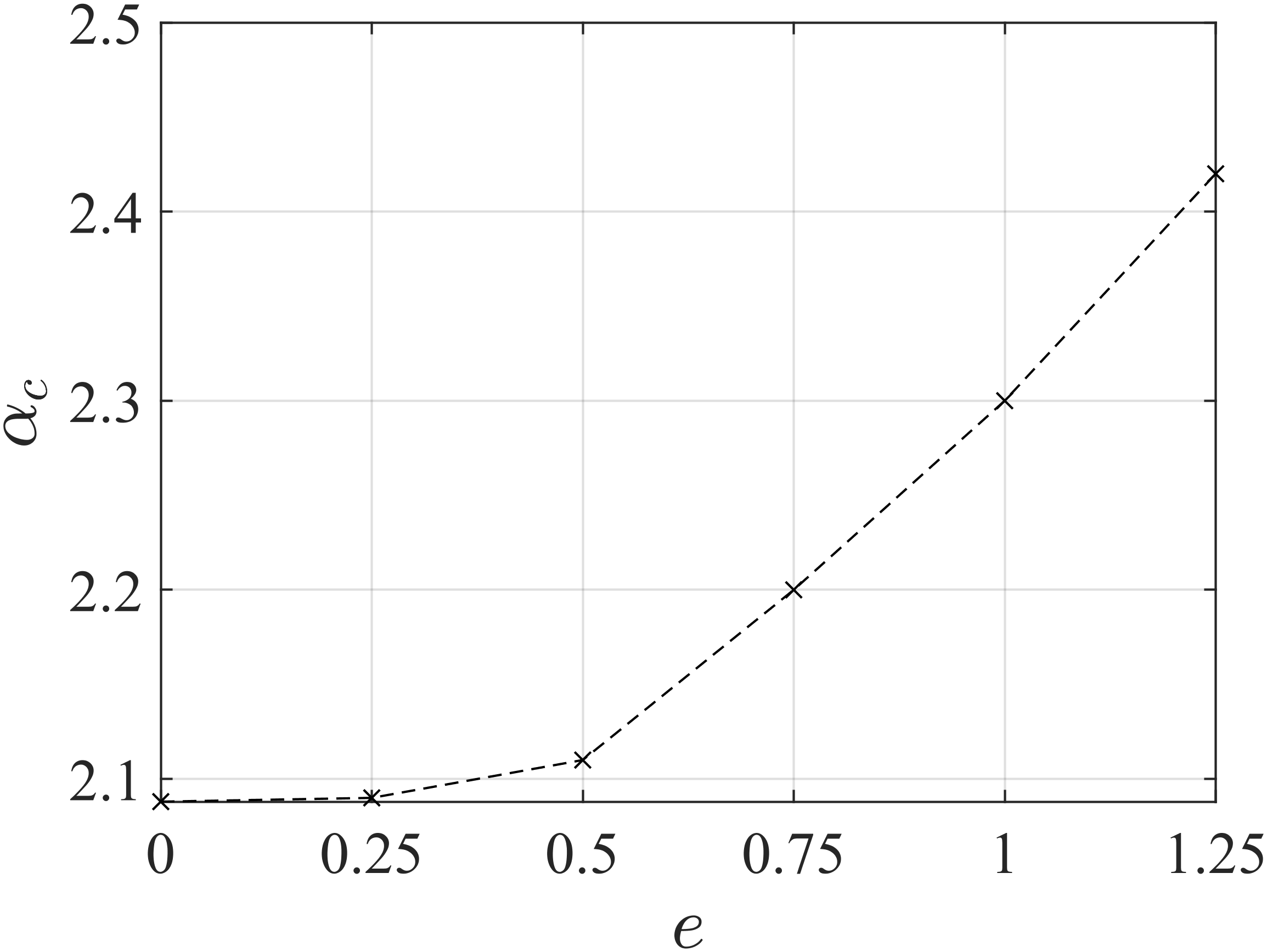}
        \caption{}
    \end{subfigure}
    \caption{a) Critical Reynolds number ($Re_c$), and b) critical wavenumber ($\alpha_c$) at various eccentricities ($e$) for the flow between the circular cylinder and an elliptical enclosure.}
    \label{fig:elliptic_critical}
\end{figure}

The Reynolds number is defined with respect to the radius of the inner cylinder as the gap varies around the azimuth. As in the concentric Couette flow, we observe that the mode shapes are also of different azimuthal wavenumbers, but cannot be defined as an exact integer due to the non-parallel nature of the flow (\cref{fig:elliptic_modes}). We observe that the most unstable mode is not an axisymmetric mode. A few unique wave modes were observed at the wide gap region of the ellipse (mode 4 and mode 5) that were observed in the secondary vortex regions of the base flow (\cite{unnikrishnan2022shear}). \Cref{fig:ur_ut_elliptic_modes} presents the normalized radial and tangential velocity profiles at different radial and angular positions. The profiles in \cref{fig:ur_ut_elliptic_modes_b} can be observed to be a combination of multiple wave modes. The modes are periodic because of the continuity of data around the chosen circle. Recently,
\citet{unnikrishnan2024taylor} performed a nonlinear direct numerical simulation of the flow with a concentric placement of the inner cylinder within an ellipse of aspect ratio 2. They had observed a hysteresis behavior in the flow indicating that there is a bifurcation at a Reynolds number around 74 from a two-dimensional base flow to the formation of Taylor-Couette cells during the forward march in $Re$, but going backward in $Re$ the transition back to two-dimensional base flow occurs at a lower $Re$ close to a value of 50. The hysteresis behavior was attributed to the residual disturbances that did not completely vanish as the $Re$ was decreased. The backward marching in Reynolds number retained some of the disturbances. In these simulations, the axial length was twice the inner radius. A Fourier spectral decomposition was used in the axial directions and the axial wavenumber imposed was restricted by the domain length to be 2.0. We therefore consider this $Re$ as the transition value for $\alpha = 2.0$.  Our present global analysis predicts that the critical Reynolds number is $48.35$ for a critical wave number $\alpha = 2.088$. Hence, this is in good agreement with the nonlinear simulations. It may be noted that the system is unstable for a lower axial wavenumber and lower Reynolds numbers compared to the concentric Couette flow.

\begin{figure}
    \centering
        \begin{subfigure}[b]{\textwidth}
        \includegraphics[width = \textwidth]{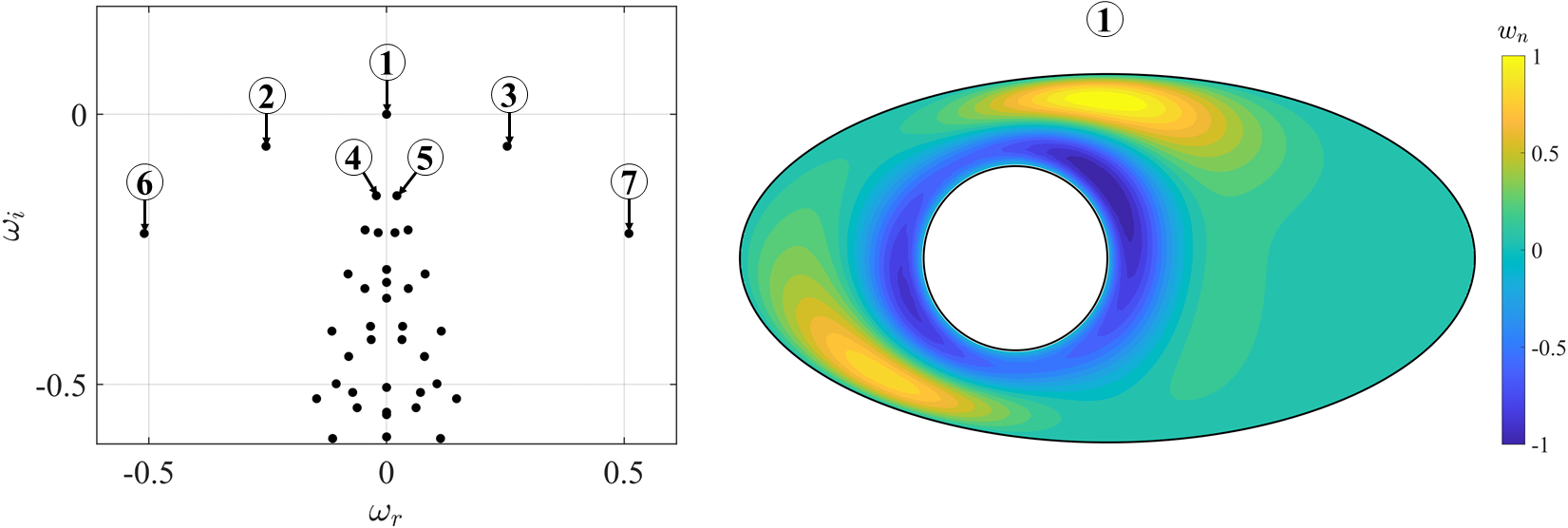}
        \caption{}
        \label{}
    \end{subfigure}
    \par \bigskip
    \begin{subfigure}[b]{0.95\textwidth}
        \includegraphics[width=\textwidth]{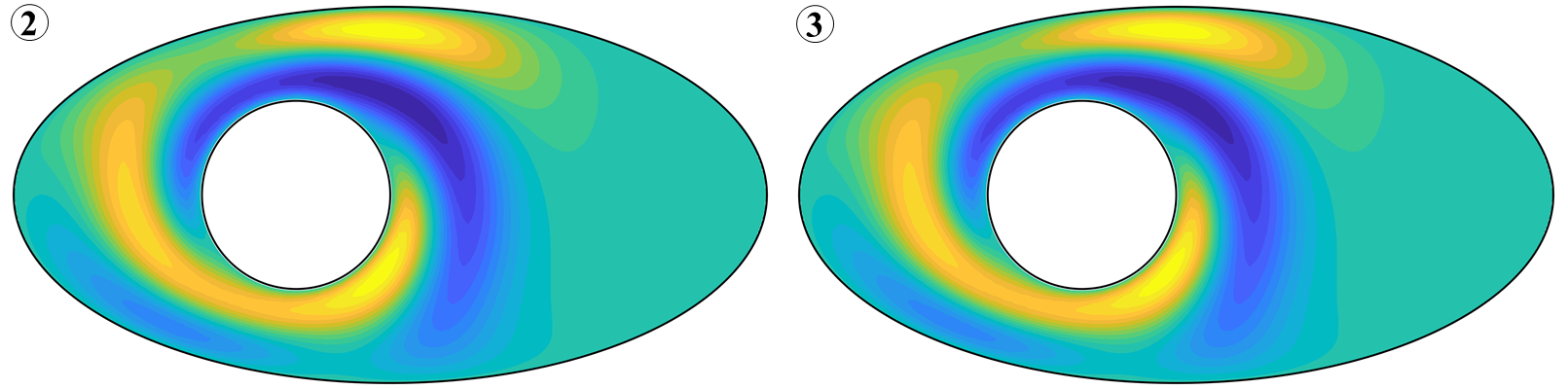}
    \end{subfigure}
    \begin{subfigure}[b]{0.95\textwidth}
        \includegraphics[width=\textwidth]{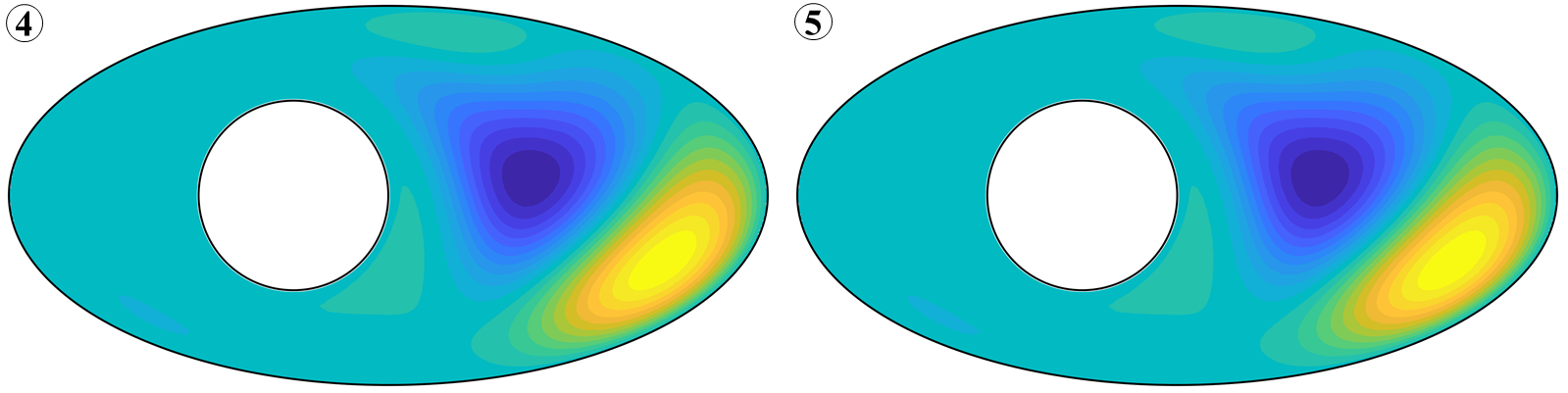}
    \end{subfigure}
    \begin{subfigure}[b]{0.95\textwidth}
        \includegraphics[width=\textwidth]{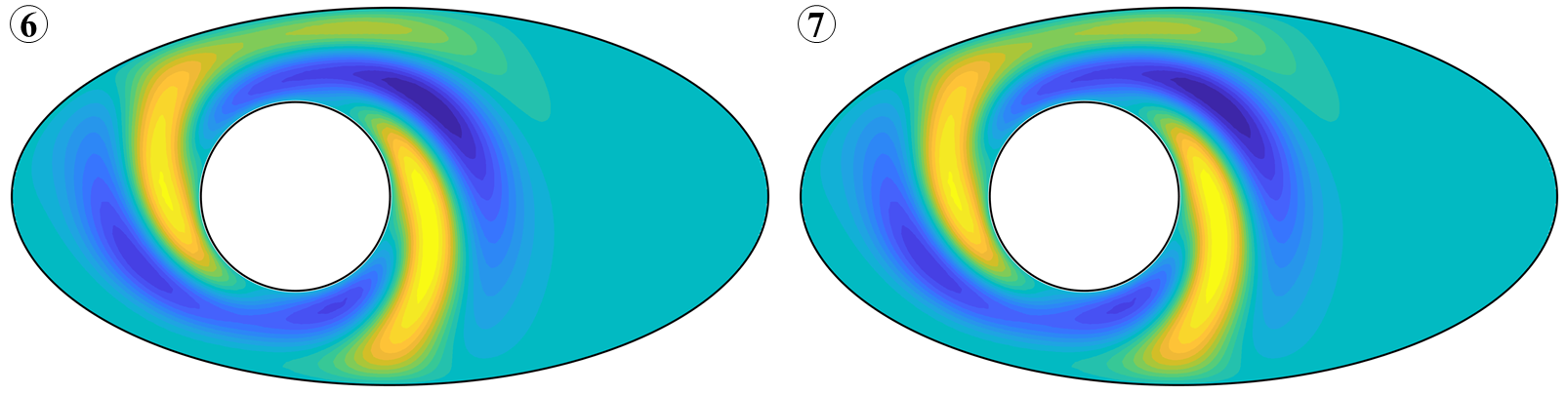}
        \caption{}
    \end{subfigure}
    \caption{a) The eigenvalue spectrum and the corresponding axial velocity profiles at the critical Reynolds number 53 and the critical wave number $\alpha = 2.3$ and b) contours of normalized axial velocity contours at $z=0$ for an eccentricity of $e=1.0$ for different modes.}
    \label{fig:elliptic_eccentric_modes}
\end{figure}


Next, we study the effects of the eccentric placement of the inner cylinder inside the ellipse. The computed critical Reynolds numbers and critical axial wavenumbers are presented in \cref{tab:critical_elliptic}. The trend of increasing critical Reynolds number and critical wave number with the increase of eccentricity observed here (\cref{fig:elliptic_critical}) is similar to that in the circular enclosure presented in \cref{fig:eccentric_Re_alpha_comparison}. The different modes at the critical $Re$ and $\alpha$ are presented in \cref{fig:elliptic_eccentric_modes}. The similarity of the shape of the spectrum and features of the modes with the concentric case is noticeable. The pairs of modes (2,3), (6,7), and (4,5) are conjugate modes and have similar mode shapes. The modes (4,5) make a pair that is dominated by waves in the wide gap region. The pairs (2,3) and (6,7) make up modes that are similar to $\beta = 1$ and $\beta =2$ modes in the concentric cylinder case. However, due to the non-parallel nature of the base flow, it is not possible to define an azimuthal wave number $\beta$.

\section{Summary}
\label{sec:conclusions}
A new global stability analysis method for non-parallel base flows in complex domains is developed using a meshless framework with PHS-RBF (Polyharmonic Spline - Radial Basis Functions) and appended polynomials. The advantage of the present method is that a linear stability analysis can now be performed on a flow within complex geometries where earlier a coordinate transformation or domain transformations with Jacobian matrices, and complex Navier Stokes operators were necessary. The present method uses a Cartesian system, which results in the same Navier-Stokes operator, irrespective of the complexity of the geometry. A significant increase in the efficiency in computations of the eigenvalue problem is achieved in comparison with other approaches. The method developed is first tested on two previously studied cases, i.e. a concentric Couette and an eccentric Couette flow. The concentric case has a parallel base flow while the eccentric case has a non-parallel base flow. The results of the two cases are validated against the published results of \cite{marcus1984simulation}, \citet{oikawa1989}, \citet{leclercq2013temporal,mittal2014finite}. Further, the generality of the global stability analysis method to compute multiple wave modes is demonstrated. The effects of the eccentric placement of the inner cylinder on the profiles of the disturbance modes are discussed. The use of the Arnoldi algorithm to compute the few important eigenmodes, along with the accuracy of the PHS-RBF method significantly decreased the computation time for the eigenmodes. The method is applied to perform global stability analysis in a complex domain with a rotating inner cylinder and an outer elliptical enclosure. The stability limits are compared with non-linear simulation results presented in \citet{unnikrishnan2024taylor}. The method is general to be applied to the study of flow stability in a variety of complex domains.

\section*{Acknowledgements}
We thank Dr. S. Shahane, previously at the University of Illinois at Urbana Champaign, for discussions during the initial stages of this study (2020) and help with the use of MemPhyS (\citet{memphys}) open-source software.

\section*{Data Availability}
The data that support the findings of this study are available from the corresponding author upon reasonable request.

\section*{Author Declarations}
The authors have no conflicts to disclose.

\appendix

\section{Assembling the matrices}
\label{appB}
The submatrices $\boldsymbol{\Phi}$ and $\boldsymbol{P}$ given in equation \cref{Eq:RBF_interp_mat_vec} can be populated as follows:
\begin{equation}
\boldsymbol{\Phi} =
\begin{bmatrix}
\phi \left(||\boldsymbol{x_1} - \boldsymbol{x_1}||_2\right) & \dots  & \phi \left(||\boldsymbol{x_1} - \boldsymbol{x_q}||_2\right) \\
\vdots & \ddots & \vdots \\
\phi \left(||\boldsymbol{x_q} - \boldsymbol{x_1}||_2\right) & \dots  & \phi \left(||\boldsymbol{x_q} - \boldsymbol{x_q}||_2\right) \\
\end{bmatrix}
\label{Eq:RBF_interp_phi}
\end{equation}
For a two-dimensional problem ($d=2$) with a maximum degree of appended polynomial set to 2 ($k=2$), there are $m=\binom{k+d}{k}=\binom{2+2}{2}=6$ polynomial terms: $[1, x, y, x^2, xy, y^2]$. Thus, the submatrix $\boldsymbol{P}$ is obtained by evaluating the polynomial terms at the $q$ cloud points.
\begin{equation}
\boldsymbol{P} =
\begin{bmatrix}
1 & x_1  & y_1 & x_1^2 & x_1 y_1 & y_1^2 \\
\vdots & \vdots & \vdots & \vdots & \vdots & \vdots \\
1 & x_q  & y_q & x_q^2 & x_q y_q & y_q^2 \\
\end{bmatrix}
\label{Eq:RBF_interp_poly}
\end{equation}

The differential operator $\mathcal{D}$, which is either one of the gradients or the Laplacian, as mentioned in \cref{sec:numerical}, can be used on these matrices. Referring to \cref{Eq:RBF_interp_mat_vec_L} we can build the matrices $\mathcal{D}[\boldsymbol{\Phi}]$ and $\mathcal{D}[\boldsymbol{P}]$ as
\begin{subequations}
\begin{equation}
\mathcal{D}[\boldsymbol{\Phi}] =
\begin{bmatrix}
\mathcal{D}[\phi \left(||\boldsymbol{x_1} - \boldsymbol{x_1}||_2\right)]_{\boldsymbol{x_1}} & \dots  & \mathcal{D}[\phi \left(||\boldsymbol{x_1} - \boldsymbol{x_q}||_2\right)]_{\boldsymbol{x_1}} \\
\vdots & \ddots & \vdots \\
\mathcal{D}[\phi \left(||\boldsymbol{x_q} - \boldsymbol{x_1}||_2\right)]_{\boldsymbol{x_q}} & \dots  & \mathcal{D}[\phi \left(||\boldsymbol{x_q} - \boldsymbol{x_q}||_2\right)]_{\boldsymbol{x_q}} \\
\end{bmatrix}
\label{Eq:RBF_interp_phi}
\end{equation}    

\begin{equation}
\mathcal{D}[\boldsymbol{P}] =
\begin{bmatrix}
\mathcal{D}[1]_{\boldsymbol{x_1}} & \mathcal{D}[x_1]_{\boldsymbol{x_1}}  & \mathcal{D}[y_1]_{\boldsymbol{x_1}} & \mathcal{D}[x_1^2]_{\boldsymbol{x_1}} & \mathcal{D}[x_1 y_1]_{\boldsymbol{x_1}} & \mathcal{D}[y_1^2]_{\boldsymbol{x_1}} \\
\vdots & \vdots & \vdots & \vdots & \vdots & \vdots \\
\mathcal{D}[1]_{\boldsymbol{x_q}} & \mathcal{D}[x_q]_{\boldsymbol{x_q}}  & \mathcal{D}[y_q]_{\boldsymbol{x_q}} & \mathcal{D}[x_q^2]_{\boldsymbol{x_q}} & \mathcal{D}[x_q y_q]_{\boldsymbol{x_q}} & \mathcal{D}[y_q^2]_{\boldsymbol{x_q}} \\
\end{bmatrix}
\label{Eq:RBF_interp_poly}
\end{equation}
\end{subequations}
In the above equations, the subscript denotes that those linear operator functions are evaluated at the corresponding cloud point.


\end{document}